\documentclass[prb,nobalancelastpage,twocolumn,nofootinbib,superscriptaddress,showpacs]{revtex4-1}

\usepackage{color,amsthm,amsmath,amsfonts,graphicx,bm}
\usepackage{bm}
\usepackage{amsfonts}
\usepackage{amssymb}
\usepackage{amsmath}
\usepackage{epsfig}
\usepackage{graphicx}
\usepackage{enumerate}
\usepackage{multirow}
\usepackage{verbatim}
\usepackage{subfigure}
\usepackage{bbold}
\usepackage{soul}
\usepackage{scrextend}

\newcommand{\tr}{\mathrm{tr}}

\newcommand{\mr}[1]{\mathrm{#1}}
\newcommand{\dg}{\dagger}

\newcommand{\mb}{\mathbf}

\newcommand{\bs}{\boldsymbol}

\newcounter{notes}
\stepcounter{notes}

\begin{document}

\title{Efficient representation of fully many-body localized systems using tensor networks}

\date{\today}

\author{Thorsten B. Wahl}
\affiliation{Rudolf Peierls Centre for Theoretical Physics, Oxford, 1 Keble Road, OX1 3NP, United Kingdom.}

\author{Arijeet Pal}
\affiliation{Rudolf Peierls Centre for Theoretical Physics, Oxford, 1 Keble Road, OX1 3NP, United Kingdom.}

\author{Steven H. Simon}
\affiliation{Rudolf Peierls Centre for Theoretical Physics, Oxford, 1 Keble Road, OX1 3NP, United Kingdom.}

\begin{abstract}
We propose a tensor network encoding the set of all eigenstates of a fully many-body localized system in one dimension. Our construction, conceptually based on the ansatz introduced in Phys. Rev. B 94, 041116(R) (2016), is built from two layers of unitary matrices which act on blocks of $\ell$ contiguous sites.  
 We argue this yields an exponential reduction in computational time and memory requirement as compared to all previous approaches for finding a representation of the complete eigenspectrum of large many-body localized systems with a given accuracy. Concretely, we optimize the unitaries by minimizing the magnitude of the commutator of the approximate integrals of motion and the Hamiltonian, which can be done in a local fashion. This further reduces the computational complexity of the tensor networks arising in the minimization process compared to previous work.
We test the accuracy of our method by comparing the approximate energy spectrum to exact diagonalization results for the random field Heisenberg model on 16 sites.  We find that the technique is highly accurate deep in the localized regime and maintains a surprising degree of accuracy in predicting certain local quantities even in the vicinity of the predicted dynamical phase transition. To demonstrate the power of our technique, we study a system of 72 sites and we are able to see clear signatures of the phase transition. Our work opens a new avenue to study properties of the many-body localization transition in large systems.

\end{abstract}

\maketitle

\section{Introduction} 

Many-body localization (MBL), a phenomenon conjectured by Anderson in 1958 for disordered, interacting quantum particles \cite{anderson1958absence}, occurs in an isolated quantum system when it fails to reach thermal equilibrium. It was shown to exist within perturbation theory for short-ranged interacting models with sufficiently strong disorder for states even at a finite energy density \cite{basko2006metal, gornyi2005interacting}. Strikingly, in one dimensional models the entire many-body spectrum can be localized \cite{pal2010mb, oganesyan2007localization}, known as full many-body localization (FMBL). As opposed to a thermalizing system where the eigenstates exhibit volume law entanglement and satisfy the eigenstate-thermalization hypothesis (ETH) \cite{deutsch1991quantum, srednicki1994chaos}, for a one-dimensional system exhibiting FMBL, all the eigenstates of the Hamiltonian are expected to obey an area law \cite{Bauer:2013jw, PalThesis}. 

The breakdown of thermalization lends itself to several interesting phenomena which are absent in a thermalizing system \cite{NandkishoreHuse_review}. Topological and symmetry breaking orders, which are destroyed by thermal fluctuations at equilibrium, can be extended to highly excited states at a finite energy density due to MBL \cite{Huse2013LPQO, Pekker2014HG, Vosk2014, Chandran2014SPT, bahri2015localization, Vosk2014}. Logarithmic growth of entanglement in the FMBL phase may allow the construction of logical qubits in an interacting system which can serve as robust quantum memories \cite{znidaric2008many, Bardason2012}. Even the quantum phase transition between the thermal and MBL phases is not described by any of the conventional theories of phase transition \cite{VHA_MBLTransition, Potter:2015ab, SondhiQPT_review, sachdev2007quantum}. Recent developments in cold atoms and various forms of synthetic quantum matter have allowed the experimental study of the phenomena of thermalization and its breakdown in a controlled manner \cite{Schreiber842, Choi1547, Smith_MBL}.

Many of the features of FMBL can be understood in terms of an extensive set of emergent quasi-local, \textit{exact} integrals of motion (qLIOM) \cite{huse2014phenomenology, serbyn2013local, ros2015integrals, chandran2015constructing}.  The phrase {\sl quasi-local} here indicates that if we trace over the operator within a region of size $x$ around where the operator is localized, we should obtain an operator proportional to the identity up to corrections that decay as $e^{-x/\xi_L}$ with \textit{localization length} $\xi_L$.   (We use the term quasi-local as compared to strictly local which would mean that outside of a sufficiently large finite sized region we obtain an operator \textit{exactly} proportional to the identity.)

A proof of the existence of qLIOMs for a strongly disordered, one-dimensional spin-$1/2$ model shows that this characteristic of the non-ergodic phase is true at least  deep in the MBL phase \cite{imbrie2016many, Imbrie2016MBL}. This emergent integrability is successful in capturing much of the phenomenology in 1D, developed based on exact diagonalization of small systems. But in the absence of numerics on sufficiently large systems or a mathematical proof, its generalization to weaker disorder or higher dimensions remains under intense investigation \cite{chandran2016higherD, Roeck2016stability}.

In the FMBL phase, the entire spectrum of the many-body Hamiltonian can be described in terms of the quantum numbers of the qLIOMs, as opposed to the case where there is a many-body mobility edge in the spectrum \cite{luitz2015many, Mondragon-Shem:2015aa, laumann2014many, kjall2014many}. As a consequence, all the eigenstates obey an area-law of entanglement\cite{Bauer:2013jw, PalThesis}, which then allows the use of highly efficient approximations involving tensor networks \cite{fannes1992, Verstraete2006, PerezGarcia2007, Schuch2008,verstraete2008matrix,Schollwock2011}. The  efficiency of these techniques is exemplified for states with area-law entanglement which can be described numerically using exponentially fewer parameters, than are required for an arbitrary state in Hilbert space. Such techniques are impressively (and provably) efficient for describing gapped ground states in one dimension\cite{Verstraete2006,Schollwock2011}.  Increasingly, similar techniques have been computationally effective  in two dimensions as well\cite{verstraete2008matrix,Orus2014}. For FMBL systems the area law holds not only for the ground state, but for the entire spectrum.

Exploiting this area-law entanglement, excited eigenstates of FMBL systems can be approximated efficiently as matrix product states (MPS) \cite{Friesdorf2015}.  Furthermore, the unitary operator diagonalizing the entire Hamiltonian can be represented as a tensor network, known as a spectral tensor network \cite{Chandran2015STN}. The algorithm to construct such a spectral tensor network proposed by Pekker and Clark does not scale efficiently with system size \cite{pekker2014encoding}. Constructing the unitary using a Wegner-Wilson flow approach also appears to be limited to small system sizes \cite{pekker2016fixed}. A proposal with an efficient scaling was given by Pollmann et al. using stacked layers of unitaries, i.e., a quantum circuit, by minimizing the fluctuations in the total energy \cite{Pollmann2016TNS}. It was suggested that the accuracy of the approximation for a given chain length can be increased by increasing the number of layers (the depth of the quantum circuit). Compared to the methods targeting eigenstates within an energy window \cite{Khemani2016MPS, Lim2016MPS, yu2015finding, serbyn2016ES}, this procedure is constructed to efficiently represent all eigenstates with sufficient accuracy, providing access to dynamical properties of local observables. 

In this work we improve upon the ansatz from Ref.~[\onlinecite{Pollmann2016TNS}] by increasing the size of the block of spins acted upon by the unitaries, while keeping the number of layers fixed at two. This corresponds to a quantum circuit of fixed depth with gates acting on several qubits. We provide analytic arguments and show numerically that this gives rise to an exponential improvement of the computational time and memory requirements. Our scheme constitutes the first scalable representation of the full set of eigenstates of FMBL systems by tensor networks: Local observables can be approximated with an error that decreases like an inverse polynomial of the computational cost.
 We use a figure of merit which is directly related to the qLIOMs and motivated by a procedure introduced by Kim et al. to identify slow operators in disorder-free non-integrable models \cite{Kim2015}. This strongly reduces the computational cost of the tensor network (TN) contractions needed to optimize our unitaries compared to using the variance as a figure of merit (as in Ref.~[\onlinecite{Pollmann2016TNS}]).  
 For concreteness, we consider the one dimensional random field Heisenberg model. We compare the numerical performance of our scheme and the one originally proposed by Ref.~[\onlinecite{Pollmann2016TNS}] even extending their study to four layers (albeit with our figure of merit to improve computational efficiency). We find our strategy to be both more accurate and computationally efficient. 

Specifically, we quantify the performance of our scheme by minimizing the commutator of the Hamiltonian with the \textit{approximate}, local integrals of motion defined through our TN ansatz. As we show, this figure of merit decomposes into strictly local parts, which allows us to evaluate it with linear cost in the system size, thus enabling us to reliably assess the performance of our ansatz in the regime where exact diagonalization is unavailable. We corroborate this by comparing the optimized TN with exact diagonalization results for 16 sites, where we observe that  the numerical value of the figure of merit indeed reflects how well the real MBL energy spectrum is approximated. We find a very high accuracy of our ansatz for unitaries acting on eight contiguous sites, and thus use the same procedure to tackle a chain with 72 sites as a function of the disorder strength. Remarkably, the ansatz fares extremely well for local observables at weak disorder and close to the MBL-to-thermal phase transition in this model. We use the fluctuations in the half-cut entanglement entropy calculated with this ansatz to estimate the location of the transition \cite{kjall2014many} which is in agreement with the exact diagonalization studies. 

In Sec. II we define the model used to perform our calculations and also highlight the phenomenological features of the FMBL phase in one dimension. In Sec. III and IV, we give a detailed description of the tensor network ansatz and the figure of merit used to diagonalize the full Hamiltonian efficiently. The numerical results and their comparison to exact diagonalization are presented in Section V. The scaling of the procedure with the total number of spins and its performance close to the MBL-thermal transition is also discussed in this section. In Sec. VI we present a summary of the results and future directions for the method.

\section{Model and its phenomenology}

We consider the canonical random-field Heisenberg model defined on a spin-$1/2$ chain \cite{pal2010mb} of $N$ sites with open boundary conditions,
\begin{align}
H = \sum_{i=1}^{N-1} \left( J \mb S_{i} \cdot \mb S_{i+1} + h_i  S_{i}^z\right) + S_{N}^z h_N^z
\label{eq:disordered_Heisenberg}
\end{align}
with $\mb S_i = \frac{1}{2} \bs \sigma_i$ and each of the $h_i^z$ is chosen from the uniform distribution bounded between $[-W,W]$, where $W$ is called the disorder strength. 
The model is known to have a dynamical phase transition into the MBL phase where all states are localized for disorder strength greater than $W_c \approx 3.5$ \cite{pal2010mb, luitz2015many}. 

In the FMBL regime the bare physical spins in the model (also known as `p-bits') can be be unitarily transformed into an extensive set of mutually-commuting quasi-local effective spins $\tau^z_i$ (also known as `l-bits') which are expected to commute \textit{exactly} with the Hamiltonian.  
\begin{align}
[H, \tau_i^z] = [\tau_i^z, \tau_j^z] = 0, \label{eq:LIOM}
\end{align}
where $\tau_i^z=U \sigma^z_i U^{\dagger}$. $U$ is the unitary operator which exactly diagonalizes the Hamiltonian. In the localized phase, the unitary transformation $U$ can be decomposed into a sequence of local unitaries so that the l-bits develop exponentially decaying tails away from site $i$ \cite{imbrie2016many}.  More mathematically, 
\begin{align}
\left\| \mr{Tr}_{i-r,i-r+1,\ldots,i+r} \left(\tau_i^z - \sigma^z_i\right)\right\|_1 \leq a \, e^{-r/\xi_L}
\label{eq:locality}
\end{align}
with positive constants $a, \xi_L$ for $N \gg r$. We use the 1-norm $\left\| A \right\|_1 = \sum_{jk} |A_{jk}|$ and consider the matrix representation of      $\tau_i^z - \sigma^z_i$ in a fixed basis.
$\xi_L$ can be defined to be the localization length of the MBL system where the trace is taken over the collection of spins within a distance $r$ of site $i$. It is important to note that the definition of the localization length is not unique. There can even be multiple localization lengths and some of them may not diverge at the MBL-thermal transition \cite{huse2014phenomenology}. 

According to Eq.~\eqref{eq:LIOM}, the Hamiltonian and the set of l-bits $\{ \tau_j^z \}$ can be simultaneously diagonalized where every eigenstate is a product state in the l-bit basis. Each eigenstate can be uniquely labelled by the eigenvalues $i_j = \pm 1$ of the set of l-bit operators $\{ \tau_j^z \}$ ($j = 1, \ldots, N$), $|\psi_{i_1 i_2 \ldots i_N}\rangle$. In the l-bit basis the Hamiltonian can be expressed in the following form, 
\begin{align}
H &= \sum_{i=1}^N J_i \tau_i^z + \sum_{i,j=1}^N J_{ij} \tau_i^z \tau_j^z + \sum_{i,j,k=1}^N J_{ijk} \tau_i^z \tau_j^z \tau_k^z \notag \\
&+ \ldots,
\end{align}
where the coefficients $J_{ijk \ldots}$ typically decay exponentially with the largest distance between two spins $|i-j|$ occurring in a particular cluster in the expansion. The probability of a coefficient being substantially larger than the typical value expected from this exponential decay, is also exponentially small\cite{imbrie2016many}. 

\section{Tensor Network Ansatz}\label{sec:spectral}

\begin{figure}
  \centering
  \includegraphics[width=0.5\textwidth]{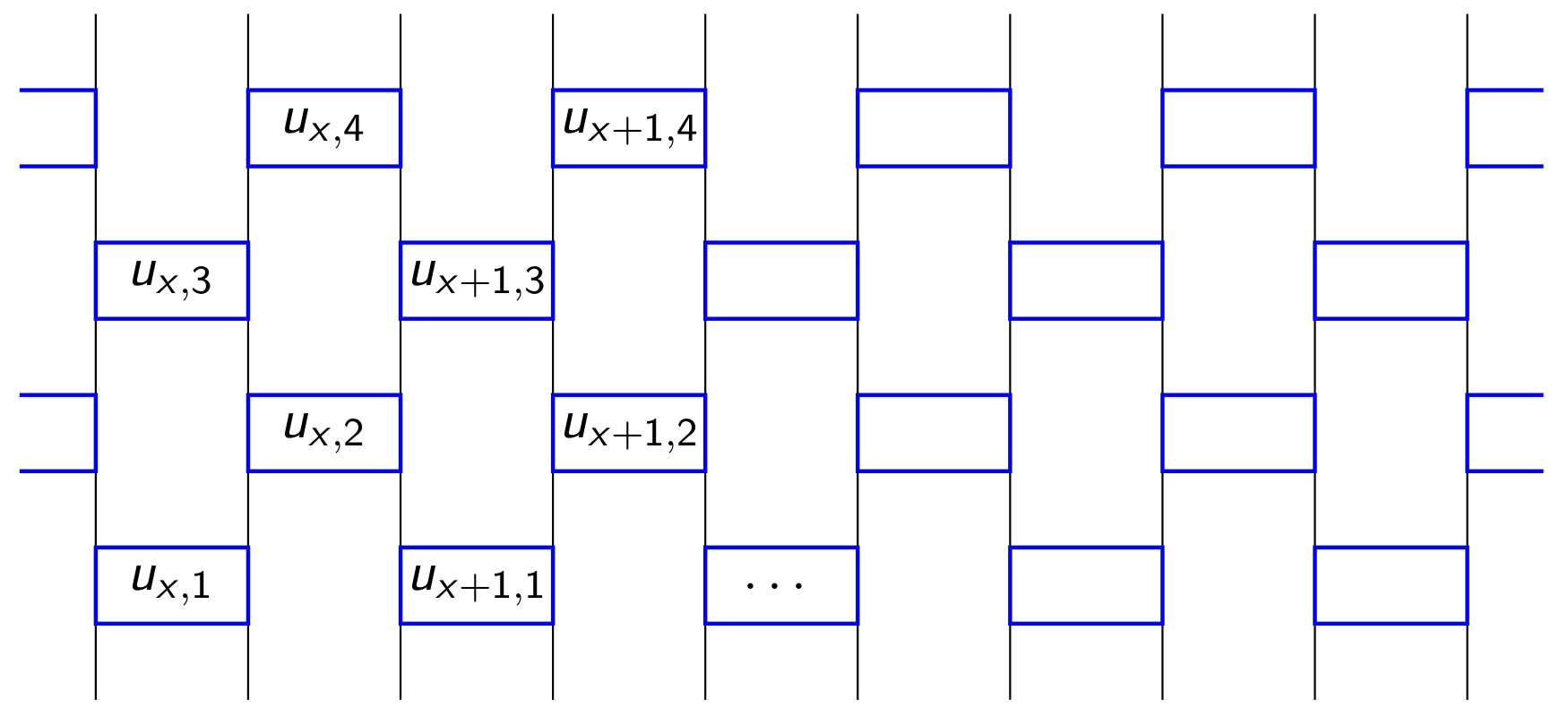}\\
  \caption{Tensor network $\tilde{U}$ as proposed in Ref.~\onlinecite{Pollmann2016TNS} with $n$ layers of $4 \times 4$ unitaries $\{u_{x,y}\}$. Since the unitary $\tilde{U}$ is supposed to approximately diagonalize the Hamiltonian, the corresponding approximate eigenstates $|\tilde \psi_{i_1,\ldots,i_N}\rangle$ 
  are the states obtained by fixing the lower open indices in the figure to be $i_1, \ldots, i_N$.  }\label{fig:Pollmann}  
\end{figure}

Tensor network states are believed to provide an efficient representation of the ground states of local gapped Hamiltonians. That is, as the system size is increased, the number of parameters required to approximate the ground state wave function with a certain fidelity (e.g., with at least 99 \% overlap) increases only polynomially with the system size. 
For MBL systems with sufficiently strong disorder (i.e., in the absence of a mobility edge), the whole spectrum of eigenstates fulfills the area law and  thus, can be efficiently represented by MPS \cite{Friesdorf2015}. However, since the number of eigenstates is exponential in the system size, for large $N$ one can only tackle the eigenstates in a certain energy window using MPS (see e.g. Refs.~[\onlinecite{Khemani2016MPS, Lim2016MPS, yu2015finding}]). On the other hand, spectral tensor networks are meant to encode an approximation to all eigenstates at once, which is a desirable property if one aims to calculate dynamical properties of local observables in MBL systems. We build on the tensor network ansatz proposed in Ref.~[\onlinecite{Pollmann2016TNS}]. It defines a unitary matrix $\tilde U$, which approximately diagonalizes the Hamiltonian, in  terms of many $4 \times 4$ unitaries, which are stacked in several layers and contracted as shown in Fig.~\ref{fig:Pollmann}. 

In the following, we argue that for this tensor network the approximation of local observables with a given accuracy 
requires the computational resources to grow superexponentially with the localization length. In contrast, for the tensor network we will suggest, they scale only exponentially with the localization length. In addition, for a fixed localization length, the error of local observables is expected to decrease as the inverse of a polynomial function in computational cost.

In Ref.~\onlinecite{Pollmann2016TNS}, the best tensor network approximation is found by minimizing the sum of the energy variances of all approximate eigenstates $|\tilde \psi_{i_1 \ldots i_N}\rangle$. The computational cost for the calculation of this quantity scales as $2^{5n}$, where $n$ is the number of layers (note that we will introduce a figure of merit below for which the minimization would only require a computational cost of order $2^{3n}$). However, it appears that $n$ needs to grow exponentially with the localization length $\xi_L$ in order to keep the accuracy of the approximation fixed on average: Within distances smaller than $\xi_L$ there are no particular restrictions on the elements of the unitary $U$ which diagonalizes the Hamiltonian. 
The number of real parameters required to  describe the unitary within that range is expected to typically scale as $2^{2 \xi_L}$. Therefore, in order to 
reproduce local observables with a given accuracy, of the order of $2^{2 \xi_L} N$ parameters are required. Since the number of parameters of the tensor network in Fig.~\ref{fig:Pollmann} is only $4^2 N \tfrac{n}{2} = 8 N n$, $n$ is required to grow as $2^{2 \xi_L}$. 
It follows that the computational resources are required to grow superexponentially with $\xi_L$ for a fixed accuracy. This makes it hard to approach the transition into the delocalized phase using the multi-layer ansatz.

\begin{figure}
  \centering
  \includegraphics[width=0.5\textwidth]{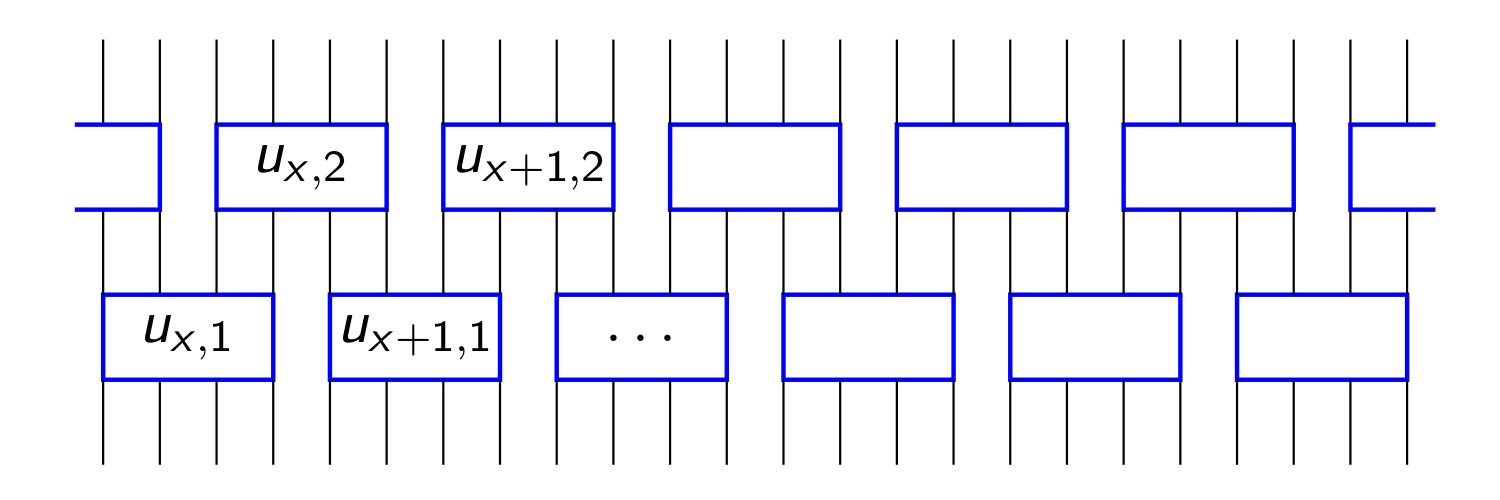}\\
  \caption{Construction of the unitary $\tilde{U}$ in terms of unitaries $u_{x,1}$, $u_{x,2}$ acting on $\ell$ sites (in this example $\ell = 4$). Again, the approximate eigenstates $|\tilde \psi_{i_1,\ldots,i_N}\rangle$ are the states obtained by fixing the lower open indices in the figure to be $i_1, \ldots, i_N$.}\label{fig:MPO}
\end{figure}

We propose to overcome this problem by increasing the range of sites acted on by the building block unitaries, instead of varying the number of layers. Thus, we stick to two layers of unitaries with $\ell$ lower and upper ``legs'' ($\ell$ is even) contracted  as shown in Fig.~\ref{fig:MPO}. Each unitary has $2^{2 \ell}$ real parameters, i.e., the total number of parameters is $2^{2 \ell+1} N/\ell$. 
Hence,  $\ell$ needs to grow only linearly with $\xi_L$ in order to keep the accuracy fixed.
The contraction cost of the tensor network arising in the variational optimization of its unitaries scales only exponentially in $\ell$ (as discussed in the following section), which is an exponential improvement over the the multi-layer ansatz. Moreover, for a fixed localization length $\xi_L$, we anticipate the error of our approximation to decrease as $\exp({-\ell/\xi_L})$ due to Eq.~\eqref{eq:locality}, 
allowing us to describe eigenstates more accurately closer to the MBL transition.
As the computational cost is exponential in $\ell$, this corresponds to a decrease in the error of local observables as the inverse of a polynomial function in computational resources, which is typical for ground states using tensor network states. On the other hand, for the multi-layer ansatz\cite{Pollmann2016TNS}, based on the above argument we expect an error of order $\exp(-2\log_2(n)/\xi_L)$, which is computationally less efficient. If one chooses both $\ell$ and $n$ larger than 2, the computational cost is approximately of order $\exp(-\ell \log_2(n)/\xi_L)$, since the above scaling argument also holds if the unitaries act on several sites within the multilayer ansatz: The number of parameters of the tensor network would increase only linearly in the number of layers and thus, reduce the error of the approximation only as the inverse of a polynomial while increasing the computational cost exponentially. Hence, keeping the number of layers fixed at two and investing computational resources only into longer unitaries fares substantially better. 

Finally, note that in both the ansatz of Ref.~[\onlinecite{Pollmann2016TNS}] and in our approach, the number of parameters required to represent the exact diagonalizing matrix $U$ with a given accuracy increases linearly with $N$ in the FMBL phase.  

\section{Figure of merit}

In order to find the unitary $\tilde{U}$ as described by our tensor network which is as close as possible to the unitary $U$ exactly diagonalizing the MBL Hamiltonian, we define a figure of merit which reflects the deviation between the two. This can be achieved by defining the approximate l-bits corresponding to $\tilde{U}$, $\tilde \tau_i^z = \tilde{U} \sigma_i^z \tilde{U}^\dg$. If they were the exact l-bits, they would commute with the Hamiltonian and with each other. The latter property is fulfilled by construction, so we define the error in our approximation as the sum of the (squared) trace norms of the individual commutator of $\tilde \tau_i^z$ with the Hamiltonian,
\begin{align}
f(\{u_{x,y}\}) &:= \frac{1}{2}\sum_{i=1}^N \tr\left([H,\tilde \tau^z_i] [H,\tilde \tau^z_i]^\dg\right) \notag \\
&= \sum_{i=1}^N \left(\tr(H^2) - \tr\left((\tilde \tau^z_i H)^2\right)\right). \label{eq:fom}
\end{align}

\begin{figure}
  \centering
  \includegraphics[width=0.48\textwidth]{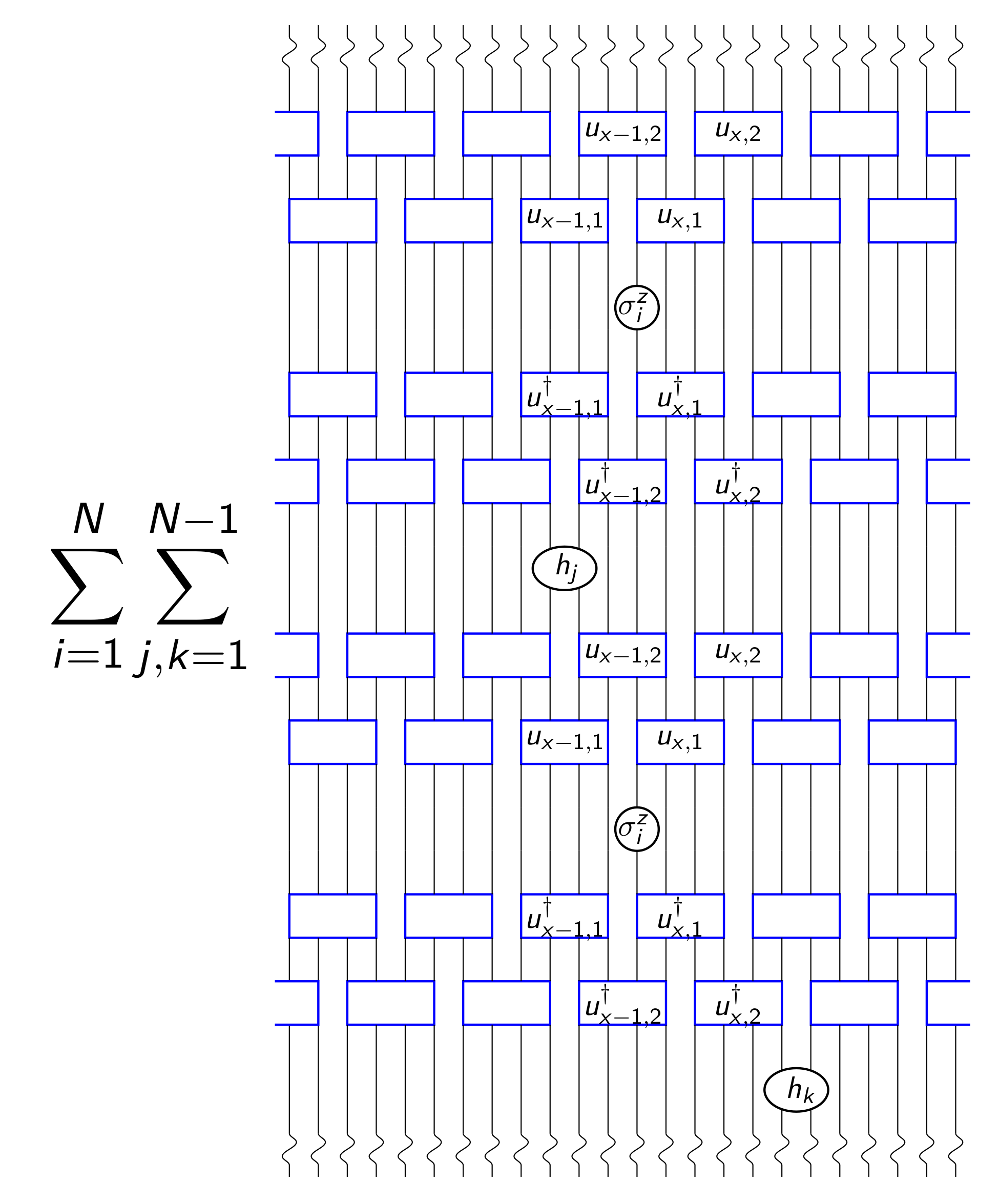}
  \caption{Sum of tensor network contractions which yields the second term, $\sum_{i=1}^N \tr\left((H \tilde \tau^z_i)^2\right)$, in Eq. \eqref{eq:fom}. The multiplications from left to right in Eq. \eqref{eq:decomposition} correspond to top to bottom in the figure. The indices of the lower wiggly lines are to be contracted with those of the corresponding upper wiggly lines. 
  For a given position $i$ of the $\sigma^z$ operator and arbitrary positions $j,k$ of the two-body Hamiltonian terms, all unitaries of the lower layer (i.e., $u_{n,1}$) apart from the ones directly connected to the $\sigma^z$ operators cancel with their adjoints and can be replaced by identities (i.e., straight vertical lines). In the example in the figure this corresponds to all unitaries $u_{n,1}$ for $n \neq x$. Furthermore, all unitaries of the second layer ($u_{n,2}$) which are not directly connected to the remaining ones of the first layer cancel and can be substituted by identities. This implies that the $x$-th summand in Eq.~\eqref{eq:decomposition} depends only on the unitaries $u_{x+1,1},u_{x,2},u_{x+1,2}$. 
The contraction corresponding to this term is shown in Fig.~\ref{fig:decomposition}. }\label{fig:fom}
\end{figure}

\begin{figure}
  \centering
  \includegraphics[width=0.48\textwidth]{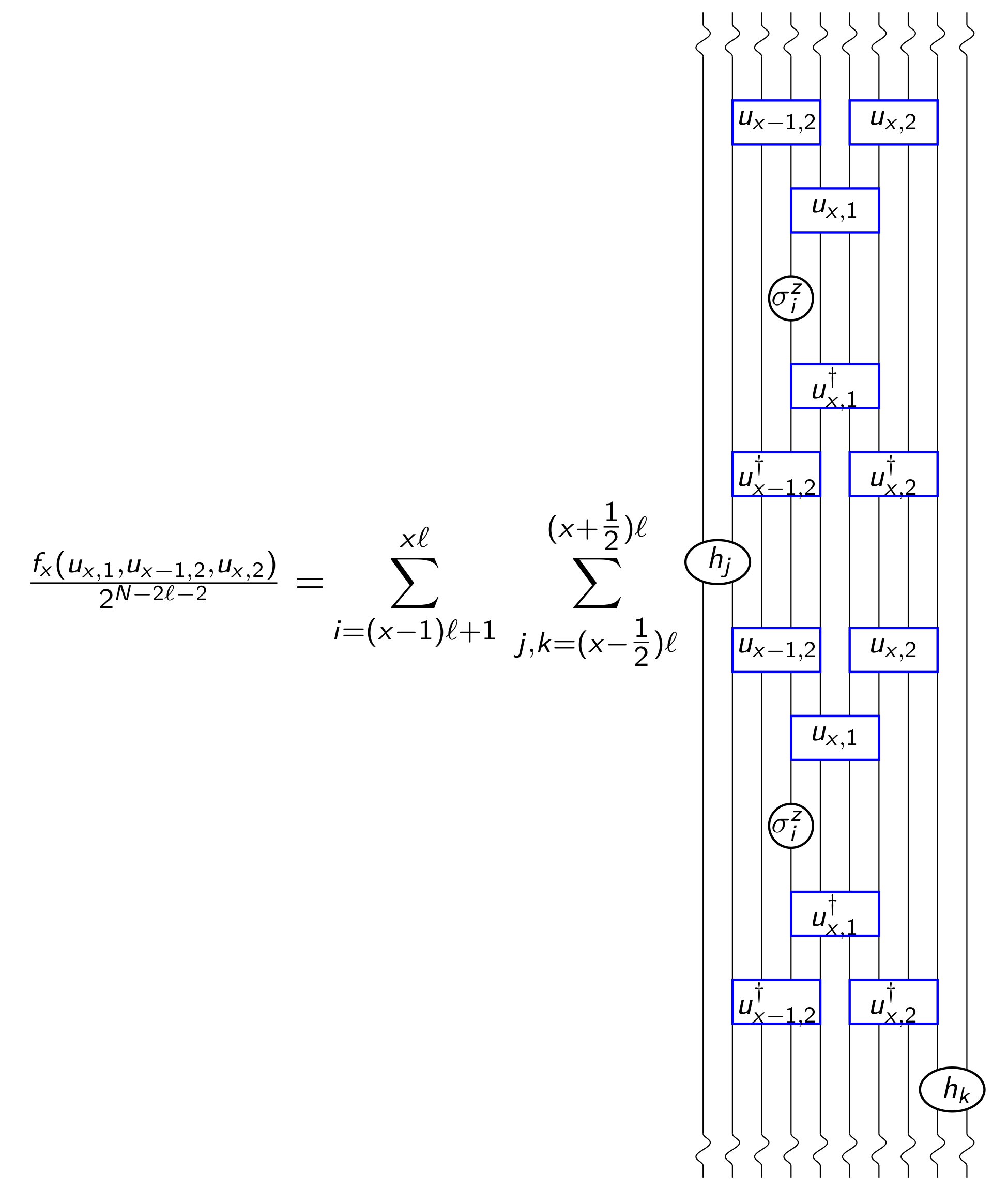}\\
  \caption{Decomposition of the figure of merit \eqref{eq:fom} into local terms resulting in Eq.~\eqref{eq:decomposition}. Again, the indices of the lower wiggly lines are to be contracted with those of the corresponding upper wiggly lines. The shown tensor network is obtained after replacing mutually cancelling unitaries in Fig.~\ref{fig:fom} by identities (vertical lines). Those forming closed loops yield a factor of 2 each, which results in the prefactor $2^{-N+2\ell+2}$ of $f_x$ shown in the figure.   
   Terms $j,k$ where $h_j$ or $h_k$ are not connected to $u_{x-1,2}$ or $u_{x,2}$ yield contributions which are independent of all unitaries $u_{x,y}$ and can thus be neglected in the local definition of our figure of merit. Note that the precise positions of $\sigma_i^z$, $h_j$ and $h_k$ depend on the indices $i,j,k$ that are being summed over, and thus the graphic depicts one example configuration. For details of our contraction scheme, see Appendix~\ref{app:decomposition}.}\label{fig:decomposition}
\end{figure}

In the following, we call $f$ the sum of the commutator norms (SCN), which will be our figure of merit. In order to minimize $f$, we evaluate the right hand side of Eq. \eqref{eq:fom}, which may naively appear exponentially hard in the number of sites $N$. However, it is possible to break it down into a sum of local terms, rendering the computational complexity linear in the system size. To that end, we first express the Hamiltonian as a sum of terms $h_i$ acting on two neighboring sites $i, i+1$, $H = \sum_{i = 1}^N h_i$. Then, the last term of Eq. \eqref{eq:fom}, $\sum_{i=1}^N \tr\left((\tilde \tau^z_i H)^2\right)$, can be easily written as a sum of tensor networks, see Fig.~\ref{fig:fom}. This term can be further decomposed into local parts as depicted in Fig.~\ref{fig:decomposition}
(using $\tilde \tau_i^z = \tilde{U} \sigma_i^z \tilde{U}^\dg$) 
\begin{align}
f(\{u_{x,y}\}) &= N \, \tr(H^2) - \sum_{i=1}^N\sum_{j,k=1}^{N-1} \tr\left(\tilde{U} \sigma_i^z \tilde{U}^\dg h_j \tilde{U} \sigma_i^z \tilde{U}^\dg h_k \right) \notag \\
&= \mr{const.} - \sum_{x=1}^{N/\ell} f_x(u_{x,1},u_{x-1,2},u_{x,2}). \label{eq:decomposition}
\end{align}
$f_x(u_{x,1},u_{x-1,2},u_{x,2})$ itself is a sum of tensor networks which only depend on $u_{x,1}$, $u_{x-1,2}$, and $u_{x,2}$, the Hamiltonian terms $h_j$ and  the $\sigma^z_i$ operators which are connected to those unitaries. Those tensor networks can be contracted by multiplying matrices of size up to $2^{\ell +1} \times 2^{\ell}$. As explained in Appendix~\ref{app:decomposition}, this results in a computational cost for the calculation of each $f_x(u_{x,1},u_{x-1,2},u_{x,2})$ which scales as $\ell^3 \, 2^{3 \ell}$, giving rise to an overall scaling of $N 2^{3 \ell}  \ell^2 $ (as there are $\tfrac{N}{\ell}$ terms $f_x$). This scaling law is a result of the figure of merit we chose. On the other hand, the minimization of the variance as in Ref.~\onlinecite{Pollmann2016TNS} requires the contraction of a matrix product operator (MPO) and scales as $L D^5 \chi^2 d^4$ (see their appendix), where $L$ is the number of tensors of the MPO, $D$ is its bond dimension, $\chi$ is the bond dimension of the MPO representing the Hamiltonian (which does not depend on the block size $\ell$) and $d$ is the physical dimension of the MPO tensors. The most efficient way to obtain such an MPO in our case is to cut each unitary vertically through the middle by performing a singular value decomposition, giving rise to a number of singular values and bond dimension of $D = 2^\ell$. Afterwards, one blocks upper and lower layers together. The physical dimension per tensor is then $d = 2^{\ell/2}$, leading to a scaling of $N \tfrac{2^{7 \ell}}{\ell}$ (since $L = \tfrac{N}{2\ell}$). Our figure of merit thus has a much lower computational complexity as $\ell$ is increased. Finally, we note that the extension of our approach to higher dimensions is still numerically efficient, since then our figure of merit still decomposes into local contributions. In contrast, calculating the variance would require the contraction of a projected entangled pair state~\cite{Orus2014}, which cannot be done exactly for large system sizes.


\section{Numerical Results}\label{sec:numerics}

\subsection{Optimization method}\label{sec:optimization}

In the following section, we will approximate the eigenstates of the Hamiltonian defined in \eqref{eq:disordered_Heisenberg}. The model possesses $U(1)$ symmetry (it conserves the total spin-$z$ component), $[H, \sum_{i=1}^N S^z_i] = 0$. Furthermore, the Hamiltonian is real in the $\sigma^z$-basis. In conventional tensor network states, symmetries of the model can be imposed on the individual tensors\cite{Sanz2009, perez2010}: Any tensor network state that is invariant under a symmetry can be written as a (possibly different) tensor network state, where all its individual tensors form a projective representation of the corresponding symmetry group, that is, they are invariant up to a phase under the action of the symmetry. In doing so, the dimensions of the tensor indices might have to be increased by a factor that is independent of the system size. The cost of variational optimization of the tensor network states usually reduces tremendously by imposing such symmetries on the tensors, as they become sparse and have fewer variational parameters. We implement a similar procedure for our spectral TN: We impose it to be real by taking all tensors as real, i.e., its unitaries are orthogonal matrices. To ensure that the total spin-$z$ component is conserved, each individual tensor $u_{x,y}$ is assumed to leave the total spin-$z$ of the block invariant, i.e. $[u_{x,y},\sum_{q = 1}^\ell \hat S^z_q]=0$, where $\hat S^z_q$ is defined in the same Hilbert space as $u_{x,y}$. Graphically speaking, this means that the sum of the spin-$z$ components on the lower legs of each tensor has to equal the sum of the spin-$z$ components of the upper legs (remember that all indices have dimension two, corresponding to spin-$1/2$ particles). All tensor entries whose indices do not fulfill this requirement are forced to be zero. This leads to a block structure of the matrix which is obtained by grouping upper and lower legs together into one single index each. Each of these blocks, say $u_B$, can be parameterized by an antisymmetric real matrix $A_B$, $u_B = e^{A_B}$, making the unitaries real and $U(1)$ symmetric.

In order to carry out the optimization, we pick initial values for the antisymmetric matrices $A_B$ parameterizing the unitaries and optimize the unitaries individually by sweeping from the left end of the chain to the right and back, until convergence is achieved. Crucially, each such minimization step requires only the evaluation of a few terms in the sum of Eq.~\eqref{eq:decomposition}. As it turns out, faster convergence is achieved by always optimizing two connected unitaries at once.

We use a quasi-Newtonian routine supplied with the gradient with respect to the parameters contained in the matrices $A_B$. This gradient comes almost for free in the contraction of the tensor network of Fig.~\ref{fig:decomposition} if one contracts its tensors in the right order, as explained in more detail in Appendix~\ref{sec:gradient}.

As it turns out, the final SCN figure of merit depends on the choice of the initial unitaries. 
For $\ell = 2$, best results are obtained by initializing the unitaries as identities and for larger $\ell$ if they are initialized according to the optimal tensor network obtained for smaller blocks. For $\ell = 4,8$, that is
\begin{align}
u^{\ell}_{x,1} &= \left(\mathbb{1} \otimes u^{\ell/2}_{2x-1,2} \otimes \mathbb{1}\right)  \left(u^{\ell/2}_{2x-1,1} \otimes u^{\ell/2}_{2x,1}\right), \\
u^{\ell}_{x,2} &= \mathbb{1} \otimes u^{\ell/2}_{2x,2} \otimes \mathbb{1},
\end{align}
where $\mathbb{1}$ is the $2^{\ell/2}$-identity matrix. 
Note that the obtained unitaries are also real and invariant under $U(1)$ symmetry. 
 For $\ell = 6$, we could only initialize the unitaries with the blocked optimal $\ell = 2$ unitaries for a given disorder realization. This corresponds simply to choosing $u^{\ell = 6}_{x,y} = u^{\ell = 2}_{3x-3+y,y} \otimes u^{\ell = 2}_{3x-2+y,y} \otimes u^{\ell = 2}_{3x-1+y,y}$.

\subsection{Comparison to exact diagonalization}\label{sec:ED}

\begin{figure}

\includegraphics[width=0.32\textwidth]{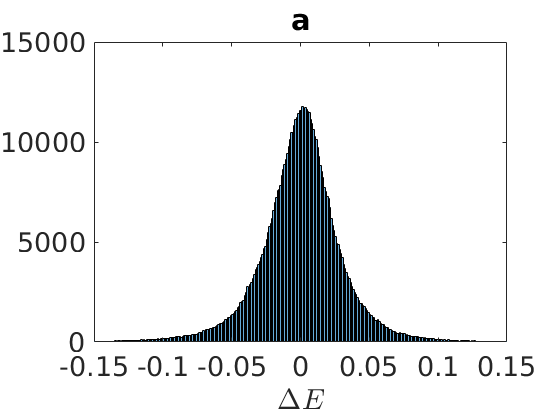}
\includegraphics[width=0.32\textwidth]{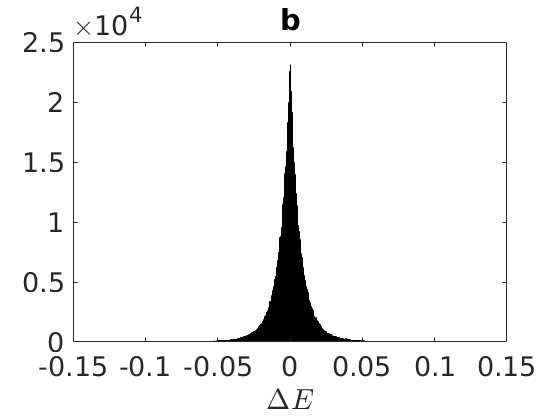}
\includegraphics[width=0.32\textwidth]{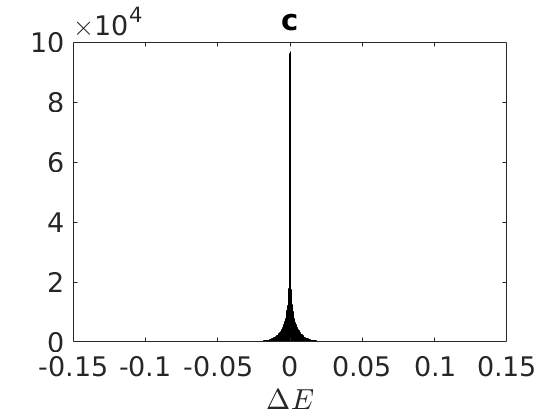}

  \caption{Comparison of the optimized tensor network $\tilde{U}$ for $\ell = 2$ (a), $\ell = 4$ (b) and $\ell = 8$ (c) with exact diagonalization for $N = 16$ and ten disorder realizations at $W = 6$. The energy differences $\Delta E$ were obtained by ordering the diagonal elements of $\tilde{U}^\dg H \tilde{U}$ in each spin-$z$ sector and subtracting them from the ordered exact eigenvalues of the Hamiltonian $H$ in the corresponding spin-$z$ sector. The plots show the concatenation of the data of all spin-$z$ sectors and disorder realizations. 
The optimized SCN figure of merit was on average $\overline{f_{\ell = 2}}/2^N = 1.011$, $\overline{f_{\ell=4}}/2^N = 0.194$, $\overline{f_{\ell=8}}/2^N = 0.0203$. 
}\label{fig:ED}
\end{figure}

\begin{figure}
\includegraphics[width=0.32\textwidth]{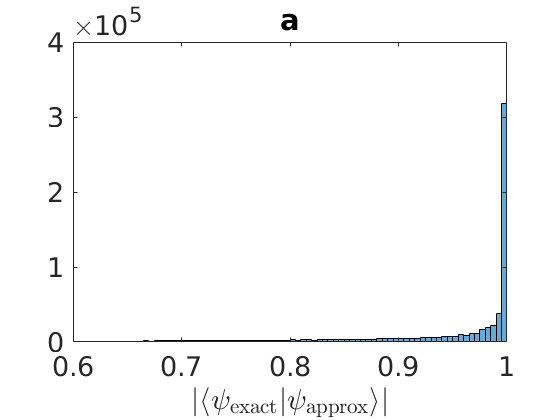}
\includegraphics[width=0.32\textwidth]{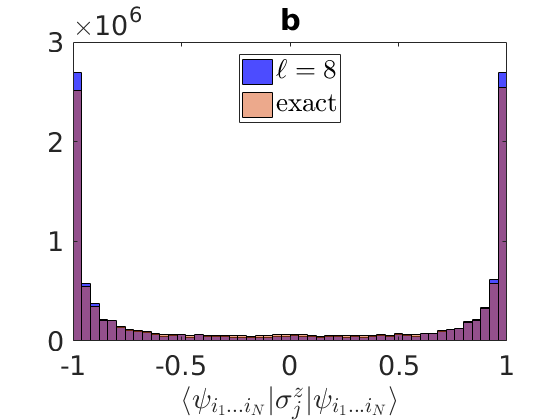}
  \caption{Comparison of the optimized tensor network $\tilde{U}$ for $\ell=8$ with exact diagonalization for $N=16$ showing the data for 10 disorder realizations at $W=6$. The approximate eigenstates are given by the columns of $\tilde{U}$. (a) Distribution of the overlap of the exact and the matched eigenstates. (b) Distribution of the expectation value of $\sigma_i^z$ over the sites and the eigenstates obtained from exact diagonalization (light brown) and the TN (blue).}    
\label{fig:eigenstates}
\end{figure}

In order to demonstrate the precision of our method for efficiently representing $U$ in the FMBL regime, we performed the optimization defined in Sec.~\ref{sec:optimization} for a system of size $N=16$ with disorder strength $W=6$ and 10 different disorder realizations using unitaries with $\ell = 2,4,8$ legs. We compare our results to the energies and the eigenstates of the Hamiltonian obtained using exact diagonalization (which was performed taking advantage of the $U(1)$ symmetry of the model). The results are shown in Figs.~\ref{fig:ED} and~\ref{fig:eigenstates}. 



In Fig.~\ref{fig:ED} the distribution of the differences between the (ordered) diagonal elements of the matrix $\tilde{U}^\dg H \tilde{U}$ and the exact energies (defined to be $\Delta E$) are plotted for the chosen values of $\ell$. The distribution narrows tremendously with increasing $\ell$ with a sharp peak at $\Delta E = 0$. 
The mean of the optimized value of the SCN figure of merit was $\overline{f_{\ell = 2}}/2^N = 1.011$, $\overline{f_{\ell=4}}/2^N = 0.194$, $\overline{f_{\ell=8}}/2^N = 0.0203$. , showing a rapid decay with $\ell$. (For an explanation of the normalization factor $2^{-N}$, see subsection \ref{sec:scaling}.) The values for the individual disorder realizations are shown in Table~\ref{table}.
The calculation time for a single disorder realization is of the order of 15 seconds for $\ell = 2$, 10 minutes for $\ell = 4$ and 4 days for $\ell = 8$ on a single CPU.
We also computed the mean variance of the approximate eigenstates (the figure of merit used in Ref. \onlinecite{Pollmann2016TNS}),
\begin{align}
\Delta H^2 = \frac{1}{2^N} \sum_{i_1 \ldots i_N} &\left(\langle \psi_{i_1 \ldots i_N} | H^2 | \psi_{i_1 \ldots i_N} \rangle \right. \notag \\
&\left.-  \langle \psi_{i_1 \ldots i_N} | H | \psi_{i_1 \ldots i_N} \rangle^2 \right). \label{eq:variance}
\end{align}
$\Delta H^2$ averaged over the different disorder realizations was $\overline{\Delta H_{\ell = 2}^2} = 0.2476$, $\overline{\Delta H_{\ell = 4}^2} = 0.0405$ and $\overline{\Delta H_{\ell = 8}^2} = 0.0035$, decaying in a very similar way as the SCN.

We find that the SCN reflects reliably the accuracy of our approximation method and thus, captures the extent to which the Hamiltonian is diagonalized by the optimal unitary matrix $\tilde{U}$. Therefore, for larger systems, where exact diagonalization is unavailable, we can use the SCN in order to assess the quality of the approximation by our tensor network. 

As a further corroboration, we computed the overlaps between the exact eigenstates and the approximate ones  for $\ell = 8$. The overlaps are in general very high, see Fig.~\ref{fig:eigenstates}(a): More than $99 \, \%$ of them have more than $60 \, \%$ overlap, with a strong peak close to an overlap of 1, which is an extremely high accuracy given that the Hilbert space dimension is $2^{16} = 65,536$. 
To show that the local properties of the eigenstates also match to high degree of accuracy, we compare the distribution over all sites of the expectation value of $\sigma_i^z$ evaluated in all the eigenstates. In Fig.~\ref{fig:eigenstates}b the distributions resulting from exact diagonalization and the spectral tensor network overlap to a remarkable precision, showing that the method has indeed converged to the eigenstates with the appropriate local features.    

\begin{figure}
\includegraphics[width=0.33\textwidth]{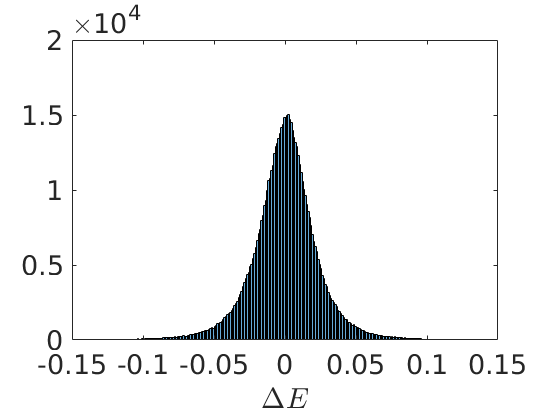}
\caption{Comparison of the optimized tensor network $\tilde{U}$ using four layers in the ansatz of Fig.~\ref{fig:Pollmann} with exact diagonalization for $N = 16$ and the same 10 disorder realizations as in Fig.~\ref{fig:ED} (i.e., $W = 6$), the unitaries of the TN being U(1)-symmetric and real. The shown energy differences have been calculated in each spin-$z$ sector separately and concatenated, also over disorder realizations. The improvement over Fig.~\ref{fig:ED}a ($\ell = 2$) is minuscule.
The average optimized SCN figure of merit was $\overline{f}/2^N = 0.747$. (For the full parameterization of the unitaries, we were not able to produce the plot due to limitations of virtual memory.)}\label{fig:n4}
\end{figure}

For comparison, we also optimized the unitaries using the ansatz in Fig.~\ref{fig:Pollmann} with four layers for the same 10 disorder realizations.  This corresponds to the scheme proposed in Ref.~[\onlinecite{Pollmann2016TNS}], extending their explicit numerical study of a network of two layers to four layers (and also using our figure of merit rather than theirs to reduce computational time). We initialized the unitaries in one series of calculations as identities and in another one with the optimized two-layer result (choosing the remaining unitaries as identities) and took the best 4-layer result in each individual case. The histogram of the error in energy is shown in Fig.~\ref{fig:n4}. 
The $U(1)$ symmetry and the fact that the unitaries are real (orthogonal matrices) imply that there is only one variational parameter per unitary.
We find $\overline{f}/2^N = 0.746$ and thus little improvement compared to the two-layer case (Fig.~\ref{fig:ED}a). The computation time per disorder realization is of the order of 20 minutes.

We also carried out such an optimization without imposing any symmetry on the unitaries, i.e., they are parameterized by an arbitrary Hermitian $4 \times 4$ matrix $H_{x,y}$, $u_{x,y} = e^{i H_{x,y}}$ with 16 variational parameters per unitary. We found that the figure of merit hardly improves over its initial value, using a standard quasi-Newtonian minimization (and minimizing four or more connected unitaries forming a column at once). This is due to the minimization algorithm being stuck in (bad) local minima, which appears to be another problem of the multi-layer approach. We managed to overcome this obstacle by the following procedure: At each step of the sweep, several completely random initial choices for the parameters of the unitaries that are being varied are optimized individually (along with their original parameters) and ranked against each other, one is able to escape these local minima and obtain much better results. By carrying out 32 of such minimizations at each sweep step, we obtained $\overline{f'}/2^N = 0.388$, which is a significant improvement over the two-layer case. 
However, the four layer calculations without imposed symmetries are much more computationally intense, requiring of the order of 1 week per disorder realization. This is of the same order of magnitude as the $\ell = 8$ calculations with two layers (note $\overline{f_{\ell=8}}/2^N = 0.0203$), while yielding worse accuracies than for $\ell = 4$ (where $\overline{f_{\ell = 4}}/2^N = 0.194$)! Hence, the potential landscape of the multi-layer ansatz, of which the algorithm has to find the global minimum, is tormented by many local minima of poor quality. As expected, we obtain much better numerical results if computational resources are invested into ``longer'' unitaries as compared to increasing the number of layers. For a comparison of the figure of merit and the variance of the individual disorder realizations, see Table.~\ref{table}.

In summary, despite our efforts, we did not succeed in making the multi-layer $\ell=2$ network calculation as accurate or computationally efficient as our multi-leg ansatz.   Thus, the numerical results corroborate our analytic arguments that the FMBL system cannot be approximated as efficiently and accurately by increasing the number of layers compared to increasing the number of legs per unitary. We also note that on relaxing the restriction of the unitaries to be real and U(1)-symmetric for four layers, a significant improvement is achieved, as the additional parameters can partially compensate for the lack of parameters that we conjectured.

\begin{center}
\begin{table}
\flushleft$\phantom{123456}\overbrace{\phantom{\ell = 2 \ \ell = 4 \ \ell = 8\ }}^{n = 2} \hspace{6pt} \overbrace{\phantom{\ell = 2 \ \ell = 4 \ \ell = 8\ \ell = 8}}^{n = 4, \ \ell = 2}$
\begin{tabular}{r| c c c | c c }
$f/2^N$ & $\ell = 2$ &$\ell = 4$&$\ell = 8$&4 layers (sym.)&4 layers \\ 
 \hline
1 &   0.98  &  0.256  &  0.0564  &  0.778  &  0.497\\
2 &   1.19  &  0.388  &  0.0053  &  0.870  &  0.478\\
3 &   0.72  &  0.106  &  0.0222  &  0.630  &  0.202\\
4 &   0.69  &  0.135  &  0.0002  &  0.467  &  0.159\\
5 &   0.58  &  0.052  &  0.0051  &  0.433  &  0.222\\
6 &   1.32  &  0.218  &  0.0209  &  0.930  &  0.615\\
7 &   0.77  &  0.038  &  0.0019  &  0.527  &  0.243\\
8 &   1.48  &  0.268  &  0.0706  &  1.274  &  0.731\\
9 &   1.65  &  0.437  &  0.0198  &  1.006  &  0.540\\
10 &   0.74 &   0.043  &  0.0010  &  0.546  &  0.196\\
 \hline
mean & 1.01 & 0.194&0.0203& 0.746&0.388  
\end{tabular}
\vspace{12pt}

\begin{tabular}{r| c c c | c}
$\Delta H$ & $\ell = 2$ &$\ell = 4$&$\ell = 8$&4 layers (sym.) \\ 
 \hline
1&    0.239 &   0.0516 &   $9.81 \cdot 10^{-3}$ & 0.186 \\
2 &   0.291 &   0.0862 &   $9.34 \cdot 10^{-4}$ & 0.208 \\
3 &   0.173 &   0.0212 &   $4.11 \cdot 10^{-3}$ & 0.149 \\
4 &   0.171 &   0.0253 &   $3.01 \cdot 10^{-5}$ & 0.116 \\
5 &   0.144 &   0.0117 &   $9.13 \cdot 10^{-4}$ & 0.106 \\
6 &   0.326 &   0.0431 &   $2.97 \cdot 10^{-3}$ & 0.223 \\
7 &   0.189 &   0.0082 &   $2.39 \cdot 10^{-4}$ & 0.128 \\
8 &   0.354 &   0.0538 &   $1.22 \cdot 10^{-2}$ & 0.302 \\
9 &   0.407 &   0.0946 &   $3.88 \cdot 10^{-3}$ & 0.231 \\
10&    0.182 &   0.0098 &   $1.94 \cdot 10^{-4}$ & 0.134 \\
 \hline
mean&    0.248 &   0.0405 &   $3.53 \cdot 10^{-3}$ & 0.178
\end{tabular}
\caption{Top: Optimized SCN figure of merit $f/2^N$ for the individual disorder realizations (column 1) and unitaries acting on $\ell = 2, 4, 8$ sites (column 2 through 4) and 4 layers of $\ell = 2$ unitaries with imposed symmetries (column 5) and without (column 6). The $\ell = 4$ ansatz performs better for all disorder realizations than the multi-layer approach. Bottom: Comparison of the variance $\Delta H$ (Eq.~\eqref{eq:variance}) for the same categories (apart from the asymmetric 4-layer case, where we were not able to compute the variance due to memory constraints). We also gather that the values of the variance and the SCN are very closely related.\label{table}}
\end{table}
\end{center}

\subsection{Scaling with the system size}\label{sec:scaling}

\begin{figure}
  \centering
  \includegraphics[width=0.52\textwidth]{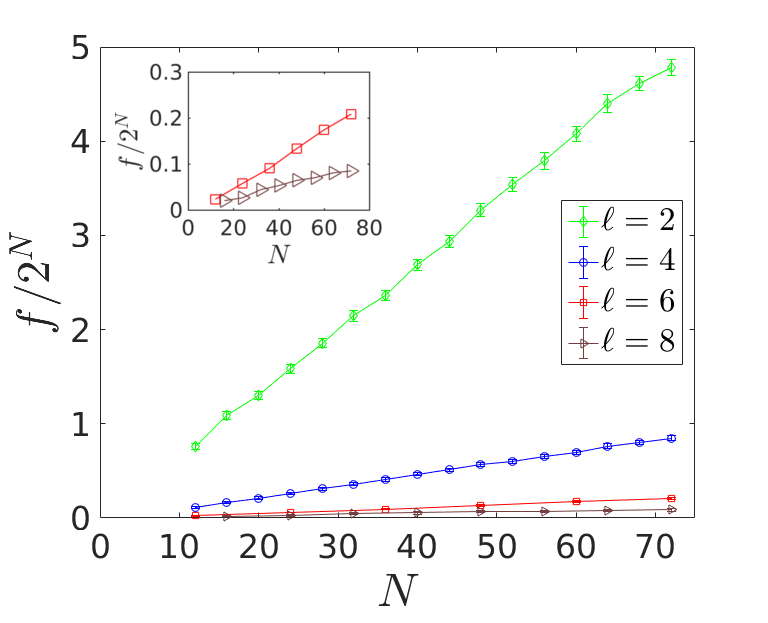}\\
  \caption{Scaling of the SCN figure of merit as a function of system size $N$ for $W = 6$ and 100 different disorder realizations optimized using unitaries of block sizes $\ell = 2,4,6$ and ten disorder realizations for $\ell = 8$. For given $N$, the same hundred (ten) disorder realizations were taken for all values of $\ell$. As discussed in the main text, we expect $f/2^N \propto N$, which is consistent with the numerical results. The error bars denote the error of the mean calculated from the distribution over disorder realizations. The inset shows an enlargement of the data for $\ell = 6$ and $\ell = 8$. There, the symbol size is at least the size of the error bars.}\label{fig:scaling}
\end{figure}

One of the primary objectives of this work is to establish our ansatz for the description of fully many-body localized systems. For a given point in the MBL phase, increasing the system size does not require an increase in $\ell$ to approximate the local properties of eigenstates with a constant accuracy (averaged over disorder realizations). Therefore, the SCN should detect a constant mismatch per lattice site and thus, increase linearly with the system size. However, recall that it is defined as a trace of an operator in the $2^N$-dimensional Hilbert space, i.e. a mismatch that is not affected by the sites far away is multiplied by the trace over the identity operator corresponding to them, which grows as $2^N$. As a result, the SCN averaged over many disorder configurations should grow as $N 2^N$. We corroborated this by optimizing our tensor network ansatz for $\ell = 2,4,6$ and $8$, and system sizes in the range between $N = 12$ and $72$ for 100 disorder realizations for $l = 2,4,6$ and 10 for $l = 8$, as shown in Fig.~\ref{fig:scaling}. We gather that on average $f/2^N$ indeed increases linearly with system size for all choices of $\ell$. 


The results in subsection \ref{sec:ED} show that the MBL eigenstates are well-represented by the optimized tensor network, i.e., by minimizing the SCN we obtain an overall unitary matrix $\tilde U$ that approximately diagonalizes the Hamiltonian to a very high accuracy for $\ell = 8$. 
The linear dependence on $N$ of the optimized SCN (divided by the dimension of the Hilbert space ($2^N$)) suggests that, in the localized region, expectation values of local observables in any eigenstate can be approximated with an error that depends only on $\ell$ and not on the size of the system. Hence, our method is able to approximate local properties of eigenstates for large system sizes, where exact diagonalization is not available. We note that while the full set of eigenstates is \textit{encoded} approximately in our tensor network, computing the eigenspectrum from it would be exponentially hard, but this is not required in practical calculations.

\subsection{Scaling with the block length}

\begin{figure}
  \centering
  \includegraphics[width=0.52\textwidth]{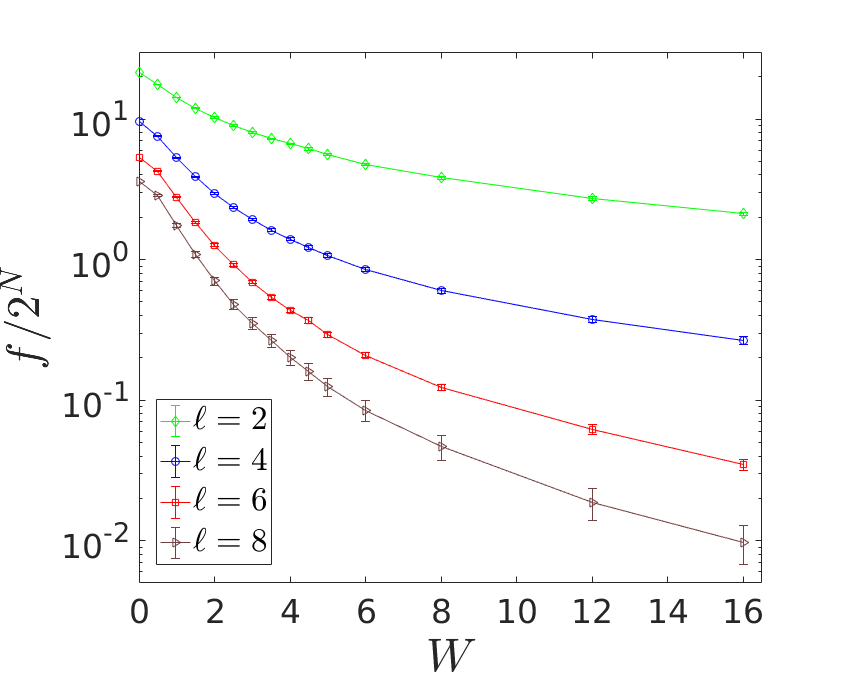}\\
  \caption{Figure of merit, SCN, for $N = 72$ as a function of disorder strength $W$ for $\ell = 2,4,6,8$. The same hundred (ten) disorder configurations were taken for $\ell = 2,4,6$ ($\ell = 8$) and all choices of $W$, adjusting only the overall prefactor of the random magnetic fields. The error bars denote the error of the mean calculated from the distribution over disorder realizations.}\label{fig:transition}
\end{figure}

\begin{figure}
\includegraphics[width=0.53\textwidth]{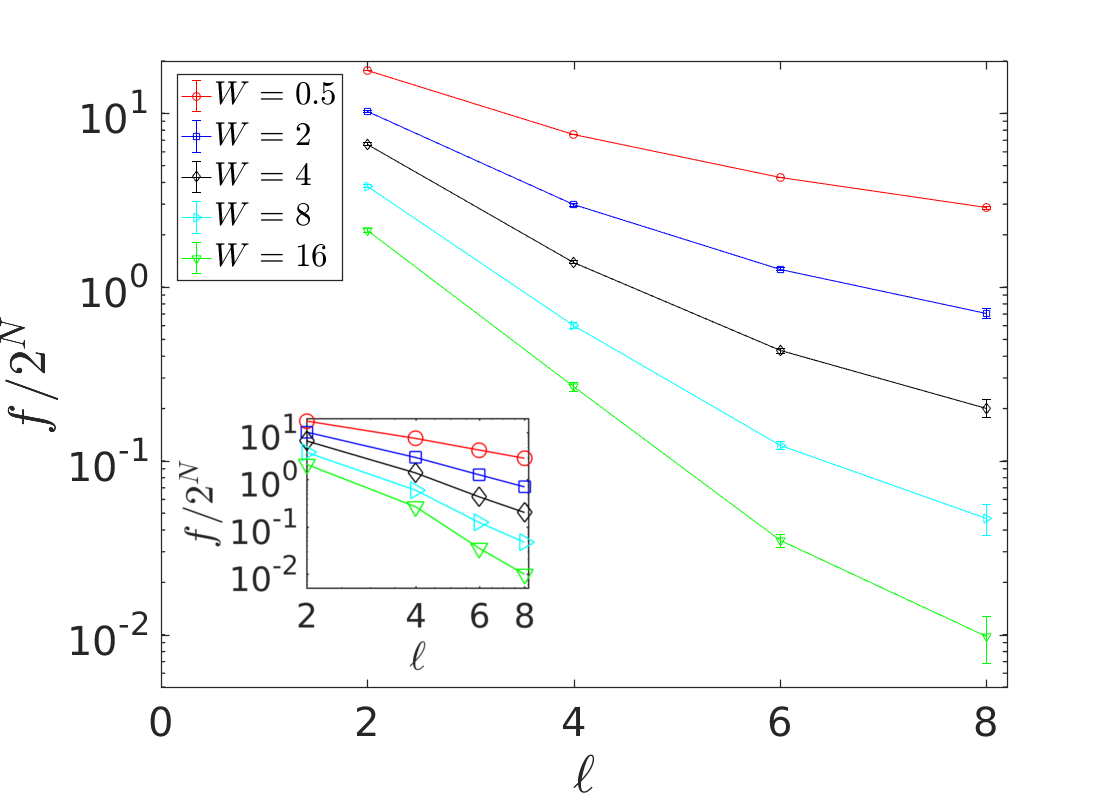}\\
  \caption{Optimized SCN averaged over disorder realizations as in Fig.~\ref{fig:transition} shown as a function of $\ell$ for various disorder strengths $W$.  The error bars denote the error of the mean calculated from the distribution over disorder realizations.
  The inset shows the same data on a log-log plot (with symbol size at least as big as the error bars). For $W = 8, 16$ the decay of the SCN is approximately exponential for $\ell = 2,4,6$ but deviates for $\ell = 8$, probably because the algorithm tends to get stuck in local minima. For $W = 0.5, 2$ the decay with $\ell$ is to a good approximation an inverse power law (exponents $-1.3$ and $-1.9$, respectively).}\label{fig:l_study}
\end{figure}

The disorder dependence of the figure of merit (scaled by $2^N$) is shown in Fig.~\ref{fig:transition} for the different values of $\ell$ at $N=72$. The improvement of the accuracy with increasing $\ell$ can be gathered from Fig.~\ref{fig:l_study}: Deep in the localized phase the relevant quantity decays almost by an order of magnitude from $\ell = 2$ to $\ell = 4$ and again from $\ell = 4$ to $\ell = 6$. For $\ell = 8$ the improvement is slightly less, presumably because our algorithm tends to get stuck in local minima. (Note that the number of parameters per unitary is 6307 for $\ell = 8$.) Nevertheless, we get by far the most accurate results for $\ell = 8$. As can be gathered from Fig.~\ref{fig:l_study}, the results point to a polynomial decay with $\ell$ in the delocalized phase ($W < 3$), whereas in the localized phase ($W > 6$) the decay with $\ell$ is exponential, as expected.  

Besides the fact that the approximation becomes worse as one approaches the transition, the SCN 
does not show any signature of the phase transition. In the following subsection, we investigate in more detail the accuracy of the eigenstates in the weakly disordered regime and the effects of the approaching phase transition into the thermal phase.

\subsection{Approaching the many-body localization transition}

\begin{figure}
\includegraphics[width=0.235\textwidth]{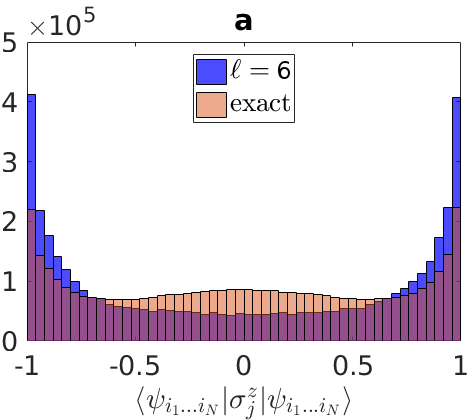}
\includegraphics[width=0.235\textwidth]{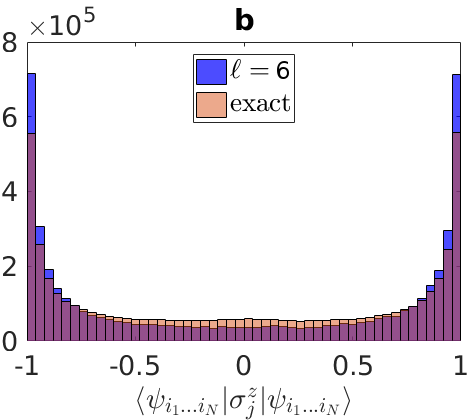}
\includegraphics[width=0.235\textwidth]{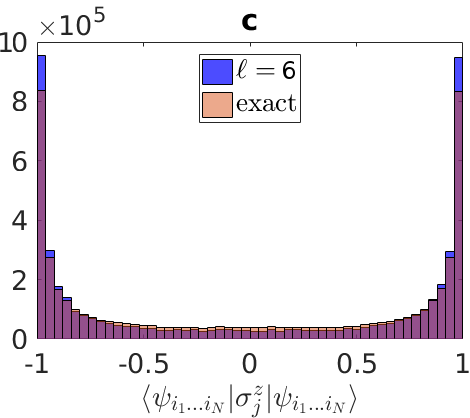}
\includegraphics[width=0.235\textwidth]{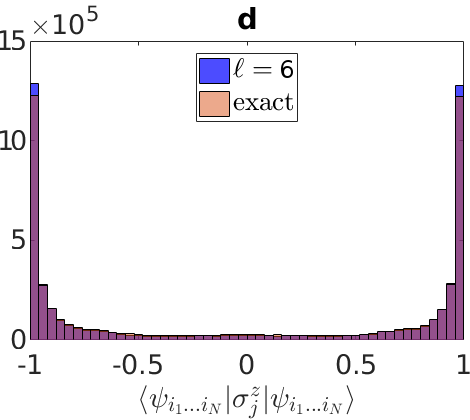}
  \caption{Comparison of the eigenstates from the optimized TN $\tilde{U}$ for $\ell=6$ and exact eigenstates from exact diagonalization for $N=12$ and 100 disorder realization at disorder strengths (a) $W=2$, (b) $W = 3$, (c) $W = 4$ and (d) $W = 6$. We present the distribution of the expectation value of $\sigma_i^z$  over the sites, eigenstates and disorder realizations from exact diagonalization (blue) and the TN (light brown).
}\label{fig:SzTransition}
\end{figure}

Deep in the MBL phase, 
existence of qLIOMs makes it amenable to approximate the eigenstates of large systems with very high accuracy using TNs. As we approach the transition into the thermal phase at weaker disorder, the eigenstates become more entangled. The ansatz with a larger number of legs is able to capture the regions of high local entanglement, allowing the method to perform appreciably well even close to the phase transition. In Fig.~\ref{fig:SzTransition} the distributions of the local observable $\sigma_i^z$ evaluated in all the eigenstates of an $N=12$ system for 100 disorder realizations, at disorder strengths $W=2$, $3$,          
 $4$ and $6$ are shown for $\ell = 6$. The average optimized SCN figure of merit $\overline{f_{\ell=6}}/2^N$ was 0.134, 0.078, 0.054 and 0.023, respectively.

\begin{figure}
\includegraphics[width=0.52\textwidth]{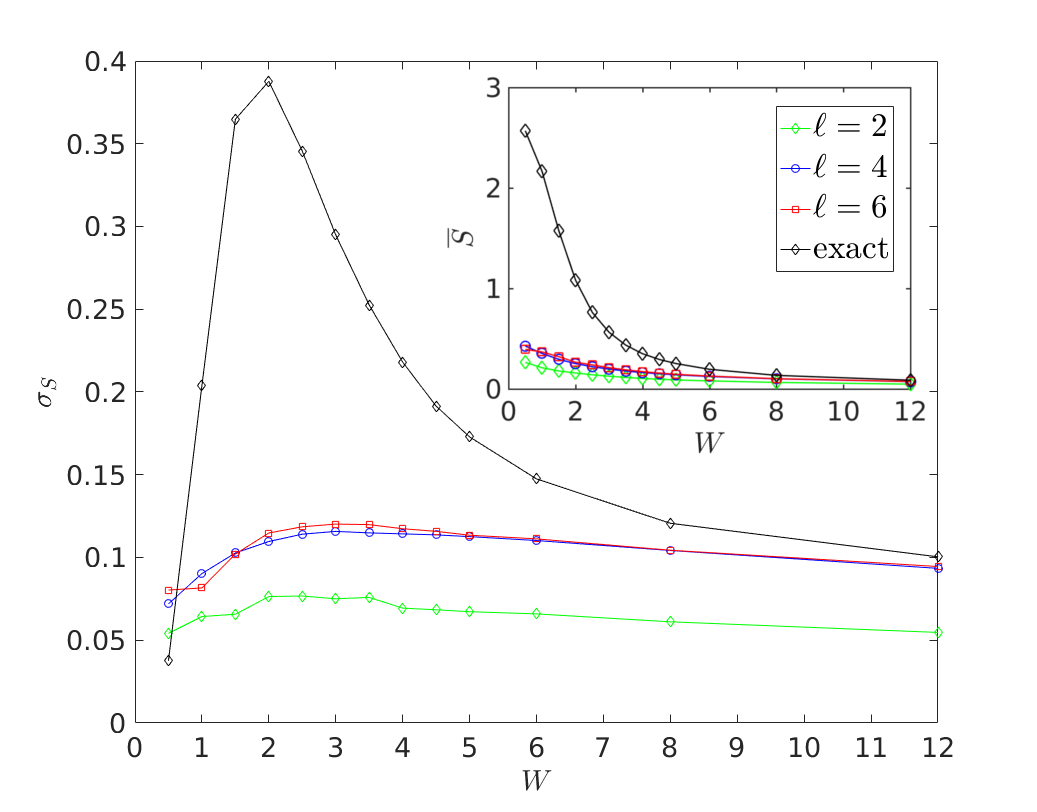}
  \caption{Standard deviation of the half-cut entanglement entropy ($\sigma_S$) for 100 disorder realizations averaged over the (approximate) eigenstates as a function of disorder strength for the TN with $\ell=2, 4, 6$, and exact diagonalization. The system size is $N=12$. The inset shows the average entanglement entropy ($\bar{S}$) for the four cases.}\label{fig:EETransition1}
\end{figure}

The distribution evaluated using the approximate eigenstates from the tensor network ansatz matches remarkably well with the exact diagonalization results in the vicinity of the MBL transition. 
The comparison with the data from exact diagonalization is good even at disorder strength $W=2$, which is expected to be on the `thermal' side of the phase transition. 
However, we cannot expect this to be the case if $N$ is increased, as opposed to the localized phase, where local observables can be reproduced with a constant accuracy for fixed $\ell$. Instead, $\ell$ would need to be scaled with $N$  to keep the accuracy fixed \cite{Pollmann2016TNS}.
In this regime the eigenstates from the tensor network ansatz have larger weight in the distribution at $\langle \sigma_i^z \rangle \approx \pm 1$, which suggests that the ansatz does not fully capture the local features of the eigenstates. The finite number of legs in the local unitaries of our tensor network ansatz enforces the qLIOMs to be always approximately conserved, but strictly local. Thus, our TN ansatz cannot resolve whether there are exactly conserved qLIOMs in the vicinity of the phase transition.

We finally turn to an extremely sensitive test of the approximate method's ability to reproduce subtle details of the MBL system. We evaluate the fluctuation in the half-cut entanglement entropy $\sigma_S$ (the standard deviation over disorder realizations of the entanglement entropy), where the entropy was averaged over all approximate eigenstates for $\ell=2,4,6$ and $N=12$ for the TN and exact diagonalization. We performed these calculations over a wide range of disorder strengths. The quantity was evaluated using 100 disorder realizations. In exact diagonalization studies, this quantity has a peak at the MBL-ETH transition which is expected to diverge with system size~\cite{kjall2014many}. Although our ansatz cannot represent any volume-law entangled states, it is expected to capture the entanglement structure at length scales of order $2 \ell$. In Fig.~\ref{fig:EETransition1} we indeed see a broad peak close to the value of disorder strength where exact diagonalization gives a relatively narrow peak. As expected, at strong disorder the exact diagonalization and the TN (for $\ell = 4,6$) tend towards the same value.

We also calculated the entanglement entropy averaged over the approximate eigenstates as given by our tensor network for system size $N = 72$ as a function of disorder strength $W$. For a specific disorder realization, the computational cost to calculate such an entropy is independent of $N$ and only depends on $\ell$: This is due to the fact the partial trace of $|\tilde \psi_{i_1 \ldots i_N}\rangle \langle \tilde \psi_{i_1 \ldots i_N}|$ gives rise to a reduced density matrix whose non-zero eigenvalues are the same as the ones of a reduced density matrix defined only on the sites which are at most one tensor block away from the entanglement cut, cf. Appendix~\ref{sec:bulk-boundary}.
This makes it possible to average over all eigenstates efficiently.

\begin{figure}
\centering\includegraphics[width=0.52\textwidth]{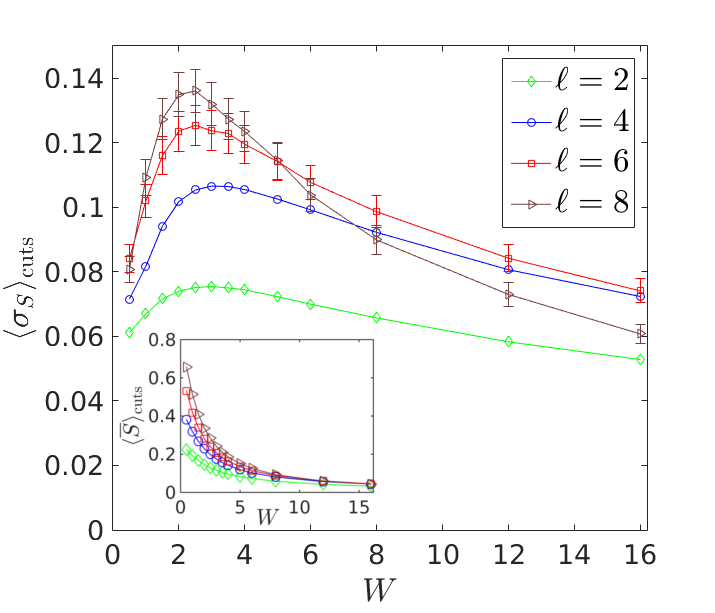}
  \caption{Standard deviation of the eigenstate-averaged half-cut entanglement entropy, followed by an average over entanglement cuts, $\langle \sigma_S \rangle_{\text{cuts}}$ (defined in the text), as a function of disorder strength. The system size is $N = 72$ and the calculations were performed on 100 disorder realizations using the TN with $\ell = 2,4,6$ and 10 disorder realizations using $\ell = 8$. For $\ell = 6$ and $\ell = 8$, the entanglement entropy was sampled over at least $1 \%$ of all eigenstates leading to an estimated relative error of about $5 \%$ (marked by error bars) of the mean and standard deviation of the entropies. The inset shows the corresponding eigenstate-averaged entanglement entropy, followed by an average over entanglement cuts ($\langle \bar{S} \rangle_{\text{cuts}}$). The errors in the inset are at most as big as the size of the symbols.}
\label{fig:EETransition2}
\end{figure}

In Fig.~\ref{fig:EETransition2} we show the statistical mean and standard deviation over 100 (10) disorder configurations of the eigenstate-averaged entanglement entropy as a function of $W$ for $\ell = 2,4,6$ ($\ell = 8$). The curves of $\langle \sigma_S \rangle_{\text{cuts}}$ and $\langle \bar{S} \rangle_{\text{cuts}}$ shown are obtained after averaging over different entanglement cuts to improve smoothness. 
The positions for the respective entanglement cuts have been chosen such that they are at least one tensor block away from the boundaries of the system. 
We observe maxima in the region $2.5 \leq W \leq 3.5$.
Averaging over entanglement cuts combined with the increased decay of $\langle \sigma_S \rangle_{\text{cuts}}$ at larger disorder strength makes the peaks much more pronounced than the $N = 12$ case.

In the insets of Figs.~\ref{fig:EETransition1} and~\ref{fig:EETransition2}, the average entanglement entropy ($\bar{S}$) increases as the disorder strength goes down but for $N=12$ this increase is still slower compared to the exact eigenstates. In the quantum critical regime, there are suggestions that the transition is driven by a subcluster of spins which are weakly entangled \cite{khemani2016critical}. With increasing system size, on the thermal side of the transition the size and the entanglement of the subcluster grows with $N$, while on the localized side of the quantum critical regime, the entanglement remains small. By varying $\ell$ and $N$ in our TN ansatz, it may be feasible to access this regime numerically which is a question suitable for future work.

\section{Conclusions and Outlook}

In this work we have made several significant advances in efficiently representing the entire set of eigenstates 
of fully many-body localized systems. Besides improving upon the tensor network ansatz proposed in Ref.~[\onlinecite{Pollmann2016TNS}], we also optimize the network by minimizing a different figure of merit (the SCN) given by the magnitude of the commutator of the Hamiltonian and the approximate qLIOMs produced by the tensor network ansatz.  This figure of merit can be evaluated by decomposing into strictly local terms leading to a much better scaling than the previous figure of merit (cf. Fig.~\ref{fig:decomposition}).

We have extended the $2$-leg, multi-layer tensor network ansatz for FMBL systems \cite{Pollmann2016TNS} to unitaries with several  legs while keeping the number of layers fixed at two.   We have shown that compared to increasing the number of layers, the extension to multiple legs  ($\ell$-legs) is far more computationally efficient --- obtaining  (exponentially) higher accuracies for the same system size and  computational cost.   

By comparing the energies and eigenstates evaluated using a TN to exact diagonalization for a chain of $16$ sites, we demonstrated that our figure of merit (SCN) reflects the accuracy of our method. In the regime where the figure of merit is small, the energy eigenvalues from the TN and exact diagonalization match extremely well. Furthermore, the distribution of expectation values of local observables in the eigenstates also matches very well with the exact diagonalization calculation. Therefore, this method is able to represent all eigenstates simultaneously to a very high degree of accuracy.

We observed that the SCN (normalized by $2^N$) increases linearly with the system size. This shows that our method only incurs a constant error per lattice site, i.e. on implementing our scheme to larger systems, for fixed $\ell$, local observables can be calculated with a size-independent accuracy.
Hence, our approximation can be readily used for large system sizes. 

In the strongly disordered regime, the error as measured by the SCN decreases exponentially with the number of legs per unitary. At weaker disorder, on approaching the MBL-ETH eigenstate transition, the local properties of the eigenstates are well approximated even close to the transition. For a large system of $N = 72$ sites, we observed peaks in the eigenstate-averaged fluctuation of the entanglement entropy for $\ell=6,8$ at  disorder strengths which are slightly lower than the critical disorder strength $W_c \approx 3.5$, which is predicted to be the critical point using exact diagonalization. 
This might indicate that exact diagonalization mildly overestimates the critical disorder strength due to finite size effects. 

The accurate construction of \textit{all} eigenstates in the FMBL phase and in the vicinity of the MBL-ETH transition for large system sizes opens the door to study several fascinating phenomena associated with the subject. As a by-product of the procedure, using our optimized unitary one directly obtains the approximate qLIOM operators in the localized phase. The ability to vary $\ell$ and study eigenstates in the vicinity of the MBL-ETH transition suggests that our procedure may be able to capture some of the scaling properties on the localized side of the quantum critical regime of the transition. Given the efficiency of the method, it may be feasible to scale the procedure to numerically address the question of many-body localization in two dimensions. Since, MBL of Floquet systems have a structure similar to that of static Hamiltonians, our method might be generalized to study the spectrum of Floquet systems exhibiting MBL as well. 
\\

\section*{Acknowledgments}

S.H.S. and T.B.W. are both supported by TOPNES, EPSRC grant number EP/I031014/1. S.H.S. is also supported by EPSRC grant EP/N01930X/1. The work of A.P. was performed in part at the Aspen Center for Physics, which is supported by National Science Foundation grant PHY-1066293. Statement of compliance with EPSRC policy framework on research data: This publication is theoretical work that does not require supporting research data.
 
\appendix

\section{Contraction scheme of our tensor network}\label{app:decomposition}

The figure of merit \eqref{eq:fom} is decomposed into local terms resulting in Eq.~\eqref{eq:decomposition}. The corresponding TN contraction is the one shown in Fig.~\ref{fig:decomposition}. Here we explain how the TN can be most efficiently contracted, resulting in a computational cost which scales like $\ell^3 2^{3\ell}$. The main idea is to block the tensors of Fig.~\ref{fig:decomposition} together such that all the new tensors correspond to matrices of size $2^\ell \times 2^\ell$ (we will see that some of them might be bigger by a factor 2). It is then possible to contract the TN from left to right or right to left while only multiplying matrices of this size. This corresponds to a computational cost of $2^{3 \ell}$ and since there are $4 \ell^3$ such terms in the sum, we obtain the claimed overall scaling.

\begin{figure*}
\flushleft
  \includegraphics[width=\textwidth]{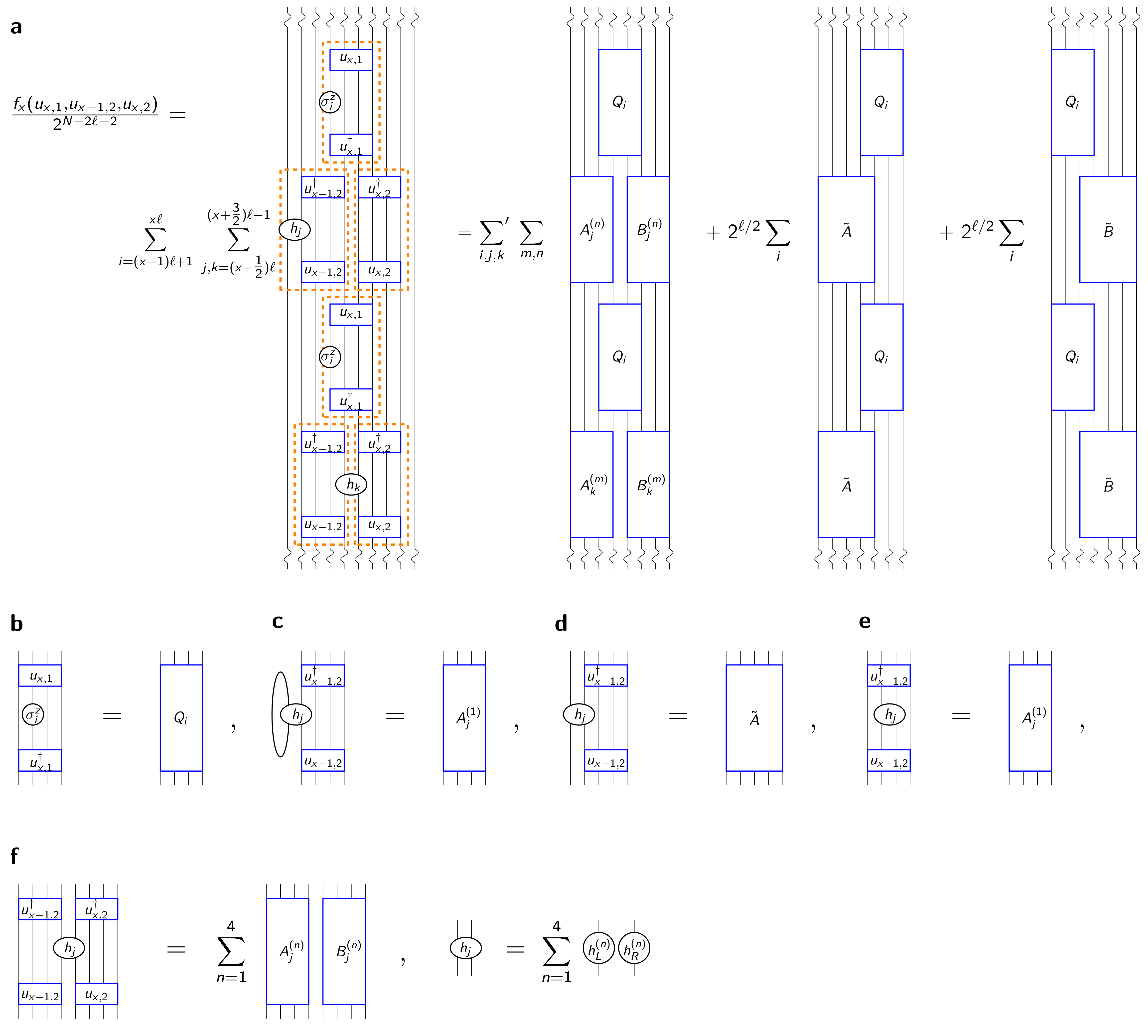}\\
  \caption{(a) The left hand side of the equation is the same as Fig.~\ref{fig:decomposition}. Here, $u_{x-1,2}$ and $u_{x,2}$ have been moved across the upper boundary coming back from the bottom due to the trace. The dashed orange boxes indicates how the tensors are blocked together in order to speed up the contraction.  
  Note that the precise positions of $\sigma_i^z$, $h_j$ and $h_k$ depend on the indices $i,j,k$ that are being summed over, i.e., the graphic on the left hand side shows only one example configuration.
  The right hand side is obtained after using the substitutions shown in (b-f). $\sum_{i,j,k}'$ is the same sum as on the left hand side with $j = k = (x-\tfrac{1}{2})\ell$ and $j = k = (x+\tfrac{3}{2})\ell$ excluded. These terms correspond to the second and third term on the right hand side, respectively. The most efficient way to contract the tensor networks on the right hand side is by contracting $A_j^{(n)}$ with $A_k^{(m)}$ ($\tilde A$ with $\tilde A$) and $B_j^{(n)}$ with $B_k^{(m)}$ ($\tilde B$ with $\tilde B$) and afterwards the resulting tensors with the $Q_i$'s. The biggest matrices to be multiplied come from the second and third term and are of size $2^\ell \times 2^{\ell + 1}$ corresponding to a cost of $2^{3 \ell + 1}$. There are $\ell$ such contractions, respectively, but to leading order $4 \ell^3$ contractions coming from the first term (multiplication of $2^\ell \times 2^\ell$ matrices, i.e., the overall computational cost is of order $\ell^3 2^{3 \ell}$.
 (b) Definition of $Q_i$. (c) Definition of $A_{j=(x-1/2)\ell}^{(n=1)}$. The upper index $(n)$ is trivial in that case. $B_{j=(x+3/2)\ell}^{(n=1)}$ is defined analogously (for the Hamiltonian term $h_{j=(x+3/2)\ell}$). (d) Definition of $\tilde A$ for $j = (x-\tfrac{1}{2})\ell$. $\tilde B$ is defined analogously. (e) Definition of $A_{j}^{(n=1)}$, where the index $(n)$ is trivial and $(x-\tfrac{1}{2})\ell < j < (x+\tfrac{1}{2})\ell$. $B_j^{(n=1)}$ is defined analogously (for $(x+\tfrac{1}{2})\ell < j < (x+\tfrac{3}{2})\ell$). For the $j$-indices we just specified for (c-e), we have $B_j^{(n=1)} = \mathbb{1}$. (f) For $j =   (x+\tfrac{1}{2})\ell$, $A_j^{(n)}$ and $B_j^{(n)}$ are obtained by using a singular value decomposition of the term $h_j$ as shown on the left and blocking $u_{x-1,2}^\dg,  h_L^{(n)},  u_{x-1,2}$ and $u_{x,2}^\dg, h_R^{(n)}, u_{x,2}$, respectively. This case is the only one where the index $(n)$ is non-trivial.
   }\label{fig:decomposition_appendix}
\end{figure*}

More specifically, we block $u_{x,1}$, $u_{x-1,2}$ and $u_{x,2}$ with its corresponding complex conjugate and the tensor between the two, respectively. This is shown in Fig.~\ref{fig:decomposition_appendix}: $u_{x,1}$, $\sigma_i^z$ and $u_{x,1}^\dg$ are contracted to form a new tensor $Q_i$. $u_{x-1,2}^\dg$ and $u_{x-1,2}$ are blocked together with whatever is in between to form $A_j^{(n)}$ and $u_{x,2}^\dg$ and $u_{x,2}$ similarly to form $B_j^{(n)}$: If $h_j$ is between $u_{x-1,2}^\dg$ and $u_{x-1,2}$ but is not connected to $u_{x,2}^\dg$ and $u_{x,2}$ as well, $u_{x-1,2}^\dg$, $h_j$ and $u_{x-1,2}$ form the new tensor $A_j^{(n=1)}$ (where the upper index is trivial). Then, $u_{x,2}^\dg$ and $u_{x,2}$ are blocked directly together, resulting in $B_j^{(n=1)} = \mathbb{1}$.
If $h_j$ is connected to all four unitaries, we split it by a singular value decomposition, $h_j = \sum_{n=1}^4 h_L^{(n)} \otimes h_R^{(n)}$. For each of the four summands, one can block $u_{x-1,2}^\dg$, $h_L^{(n)}$ and $u_{x-1,2}$ to form $A_j^{(n)}$ and $u_{x,2}^\dg$, $h_R^{(n)}$ and $u_{x,2}^\dg$ to form $B_j^{(n)}$, respectively. Finally, if $h_j$ is only connected to $u_{x,2}^\dg$ and $u_{x,2}$, they together are blocked to $B_j^{(n=1)}$, whereas $u_{x-1,2}^\dg$ and $u_{x-1,2}$ are blocked directly together, resulting in $A^{(n=1)}_j = \mathbb{1}$. The lower blocks $A_k^{(m)}$ and $B_k^{(m)}$ in Fig.~\ref{fig:decomposition_appendix} are defined in the same way. The only two exceptions are $j = k = (x-\tfrac{1}{2})\ell$ and $j = k = (x+ \tfrac{3}{2})\ell$. In the earlier case, $A_j^{(1)} = A_k^{(1)}$ obtains two additional indices (an extra upper and lower leg in the graphical representation), which is why we call it $\tilde A$. $B_j^{(1)} = B_k^{(1)} = \mathbb{1}$ in that case. For $j = k = (x+ \tfrac{3}{2})\ell$, the assignment is the other way around, resulting in the big tensor $\tilde B$.


\section{Calculation of the gradient}\label{sec:gradient}

As pointed out in subsection \ref{sec:optimization}, due to the presence of $U(1)$ symmetry and the fact that the Hamiltonian is real (in the $\sigma^z$-basis), we parameterize the unitaries in terms of real antisymmetric matrices $A_B$ corresponding to the blocks $B$ of conserved $U(1)$ charge. For $\ell \geq 4$ the optimization gets tremendously sped up by providing the gradient of the function to be minimized. Hence, the derivative is, calling $\{a^m_{x,y}\}_m$ the parameters contained in  all blocks $A_B$ corresponding to a certain unitary $u_{x,y}$, 
\begin{align}
\frac{\partial f(\{u\})}{\partial a^m_{x,y}} = \begin{cases}
- \frac{\partial f_x(u_{x,1},u_{x-1,2},u_{x,2})}{\partial a^m_{x,y}}, \ \ \ \mr{if} \ y = 1 \\
- \frac{\partial f_x(u_{x,1},u_{x-1,2},u_{x,2})}{\partial a^m_{x,y}}
- \frac{\partial f_{x+1}(u_{x+1,1},u_{x,2},u_{x+1,2})}{\partial a^m_{x,y}}, \\ \hspace{0.2\textwidth} \mr{if} \ y = 2.
\end{cases}
\end{align}
\begin{figure}
  \centering
  \includegraphics[width=0.35\textwidth]{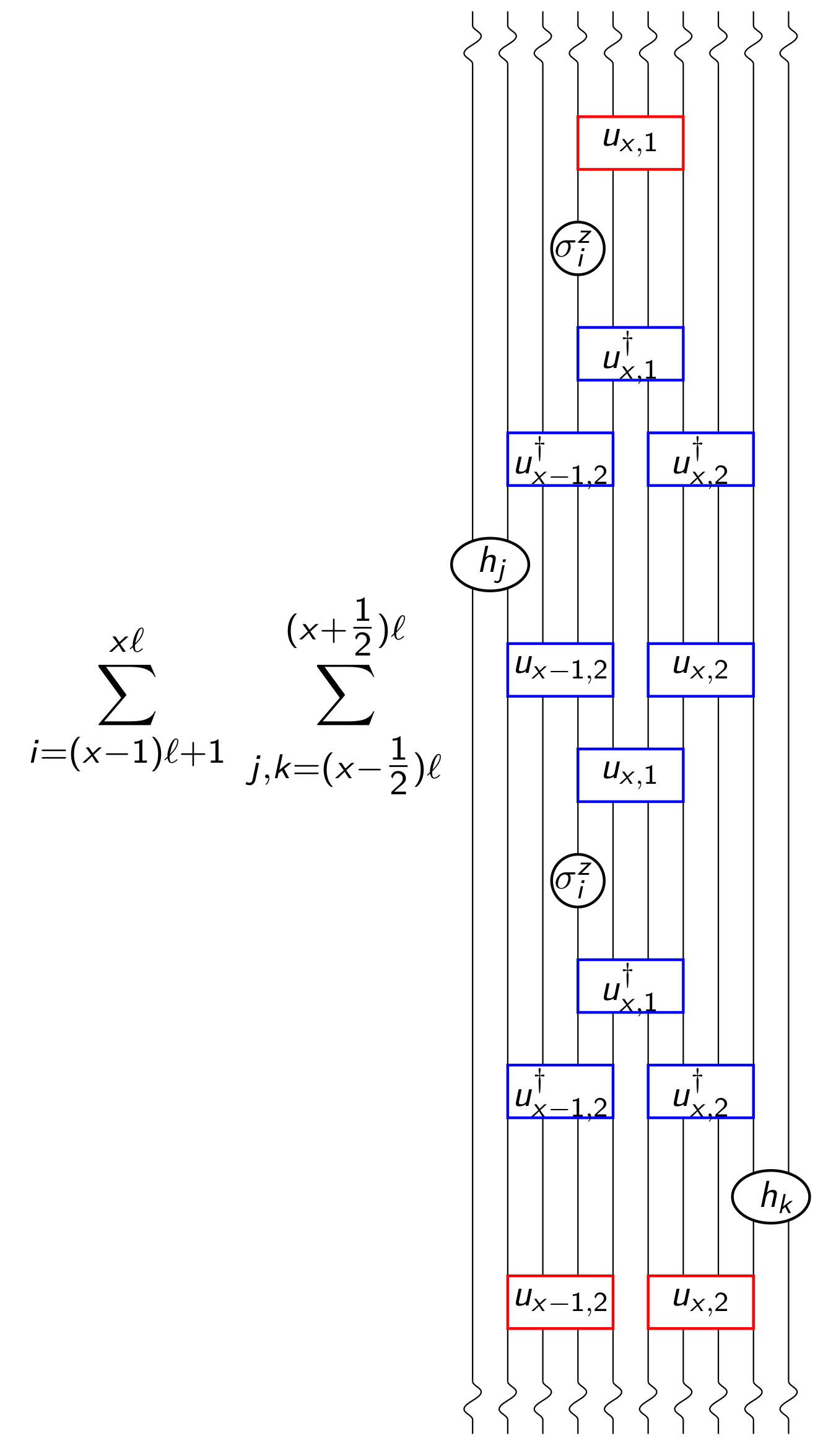}\\
  \caption{The shown tensor network contraction results in  $2^{-N+2l+2} M$, with the $2^\ell \times 2^\ell$ matrix $M$ described in the main text if the unitary of which the derivative is being taken, is cut out on the very top or very bottom (marked in red). 
   The tensor network can be contracted with the same computational scaling as before ($\ell^3 2^{3\ell}$) by using the same blocking as in Fig.~\ref{fig:decomposition_appendix}. The resulting blocks $A_j^{(n)}$, $B_k^{(m)}$ etc. are contracted in such an order that the one which contains the unitary to be varied is contracted with last. E.g., if $u_{x,1}$ is being varied (which is contained in the upper $Q_i$), one contracts first $A_{j}^{(n)}$ with $A_{k}^{(m)}$ and the resulting tensor with the lower $Q_i$ and subsequently with the contraction of $B_{j}^{(n)}$ and $B_{k}^{(m)}$.\label{fig:derivative}}
\end{figure}
In order to evaluate the derivatives on the right hand side, we contract the local tensor network of Fig.~\ref{fig:decomposition} as shown in Fig.~\ref{fig:derivative} and cut out the tensor the derivative is taken of at the very top or the very bottom, respectively, before taking the overall trace indicated by wiggly lines. The contraction is again most efficiently carried out using the same blocking as in Fig.~\ref{fig:decomposition_appendix} (where the block from which the unitary has been taken has to be modified).
Since we cut out a tensor with $\ell$ lower and upper legs, the result of the contraction is a $2^\ell \times 2^\ell$ matrix, say $M$. This matrix can be used to obtain both $f_x$ by putting back the missing tensor, $f_x(u_{x,1},u_{x-1},u_{x,2}) = \tr(M u_{x,y})$, and the desired derivative, $\frac{\partial f_x(u_{x,1},u_{x-1},u_{x,2})}{\partial a^m_{x,y}} = 4 \, \mathrm{Re} \left(\tr(M \frac{\partial u_{x,y}}{\partial a^m_{x,y}})\right)$ without the need for any additional contractions.

\section{Calculation of the entanglement entropy for large systems}\label{sec:bulk-boundary}

\begin{figure*}
  \centering
  \includegraphics[width=\textwidth]{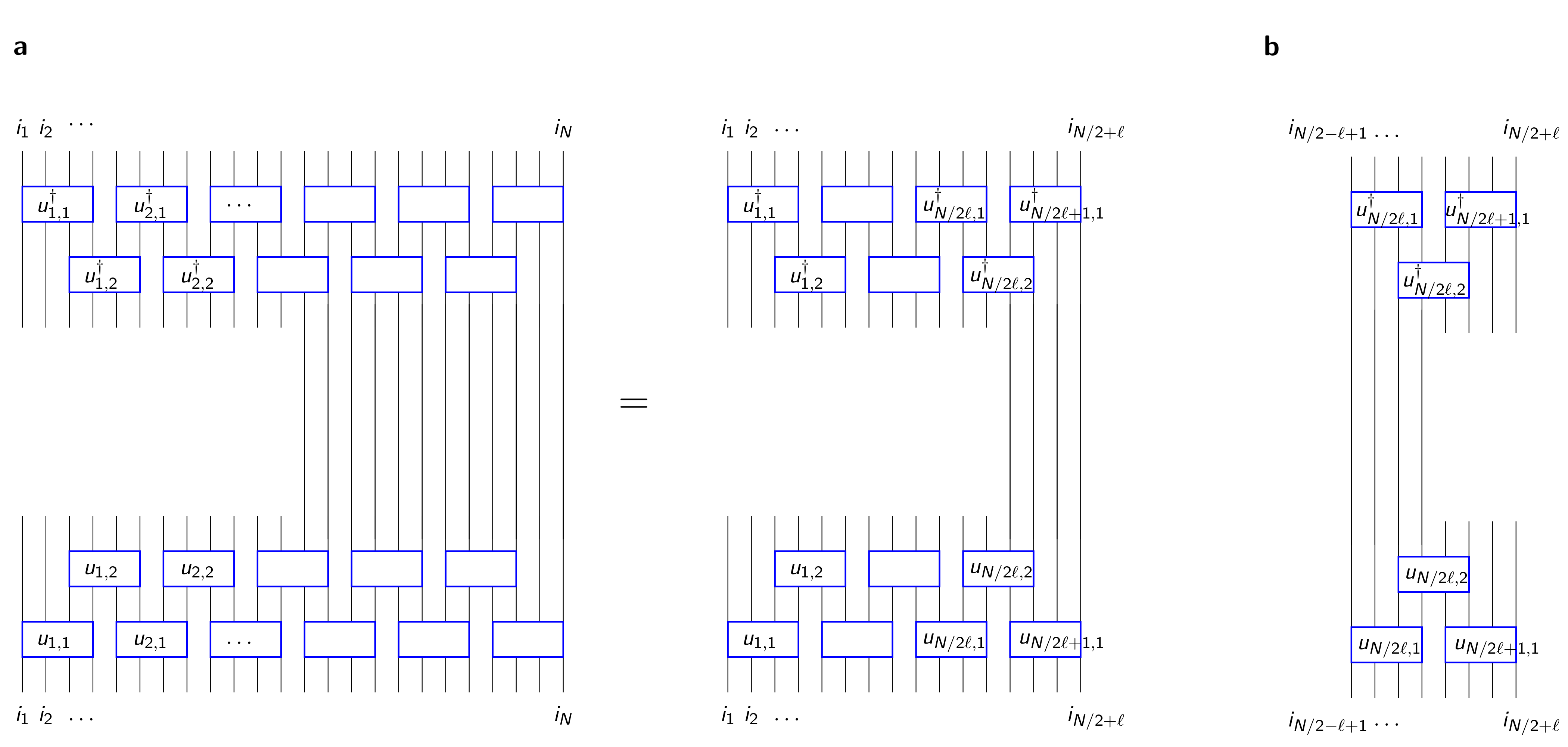}\\
  \caption{(a) The left hand side of the equation is the reduced density matrix $\rho$ obtained by tracing out sites $\tfrac{N}{2}+1, \ldots, N$ according to Eq.~\eqref{eq:rho_half}. By cancelling unitaries and their adjoints, the right hand side is obtained, cf. Eq.~\eqref{eq:rho_less}. As explained in the text, the non-trivial spectrum of $\rho$ is the same as the one of the reduced density matrix $\sigma$ defined in (b).}\label{fig:entropies}
\end{figure*}

The von Neumann entropy of an approximate eigenstate $|\tilde \psi_{i_1 \ldots i_N}\rangle$ for an entanglement cut through the middle of the system is defined via its reduced density matrix
\begin{align}
\rho = \tr_{{N/2+1}, \ldots, N} \left(|\tilde \psi_{i_1 \ldots i_N}\rangle \langle \tilde \psi_{i_1 \ldots i_N\rangle}|\right) \label{eq:rho_half}
\end{align}
and given by $S = -\tr(\rho \ln(\rho))$. $\rho$ corresponds to the tensor network contraction shown in Fig.~\ref{fig:entropies}a and can be simplified by replacing unitaries which are contracted with their adjoints by identities. We obtain a representation for $\rho$ in terms of the unitaries $\{u_{1,1}, u_{1,2}, \ldots, u_{N/2\ell,1}, u_{N/2\ell,2}, u_{N/2\ell+1,1}\}$. They define a new tensor network, which in turn defines a state $|\phi_{i_1, \ldots i_{N/2+\ell}}\rangle$, which is independent of the indices $i_{N/2+\ell+1}, \ldots, i_N$. Therefore, we can write
\begin{align}
\rho = \tr_{{N/2+1}, \ldots, {N/2+\ell}} \left(|\phi_{i_1 \ldots i_{N/2+\ell}}\rangle \langle \phi_{i_1 \ldots i_{N/2+\ell}}|\right). \label{eq:rho_less}
\end{align}
Calculating the von Neumann entropy of $\rho$ directly would still be exponentially hard in $N/2$. However, the von Neumann entropy depends only on the eigenvalues $\lambda_m$ ($m \in \mathbb{N}$) of $\rho$. If we carry out a Schmidt decomposition of $|\phi_{i_1 \ldots i_{N/2+\ell}}\rangle$ across the cut between site $N/2$ and $N/2+1$, 
\begin{align}
|\phi_{i_1 \ldots i_{N/2+1}}\rangle = \sum_{m} \mu_m |\phi_L^{(m)} \rangle \otimes |\phi_R^{(m)}\rangle,
\end{align}
the desired eigenvalues are $\lambda_m = \mu_m^2$ ($\mu_m > 0$), since
\begin{align}
\rho = \sum_m \mu_m^2 |\phi_L^{(m)}\rangle \langle \phi_L^{(m)}|.
\end{align}
The crux is that this has the same non-trivial spectrum as the reduced density matrix\cite{bulk-boundary}
\begin{align}
\sigma &= \sum_m \mu_m^2 |\phi_R^{(m)}\rangle \langle \phi_R^{(m)}| \notag \\
 &= \tr_{1, \ldots, N/2} \left(|\phi_{i_{1} \ldots i_{N/2+\ell}}\rangle \langle \phi_{i_{1}\ldots i_{N/2+\ell}}|\right)  \notag \\ 
 &= \tr_{N/2-\ell+1, \ldots, N/2} \left(|\varphi_{i_{N/2-\ell+1} \ldots i_{N/2+\ell}}\rangle \langle \varphi_{i_{N/2-\ell+1}\ldots i_{N/2+\ell}}|\right),
\end{align}
where $|\varphi_{i_{N/2-\ell+1} \ldots i_{N/2+\ell}}\rangle$ is defined only in terms of the unitaries $u_{N/2\ell,1}, u_{N/2\ell,2}, u_{N/2\ell+1,1}$. Hence, $S = -\tr(\rho \ln(\rho)) = -\tr(\sigma \ln(\sigma))$. $S$ depends only on the indices $i_{N/2-\ell+1}, \ldots, i_{N/2+\ell}$ and can be evaluated efficiently for large $N$ by calculating the reduced density matrix $\sigma$ shown in Fig.~\ref{fig:entropies}b. The treatment of entanglement cuts at other positions is analogous.

\bibliography{biblioMBL}{}

\end{document}